%% file: vz_xbb_combo_note.tex
\documentclass[prd,amsmath,amssymb]{revtex4}

\usepackage{graphicx}
\usepackage{dcolumn}
\usepackage{bm}
\def\MET{{\mbox{$E\kern-0.57em\raise0.19ex\hbox{/}_{T}$}}}
\def\met{{\mbox{$E\kern-0.57em\raise0.19ex\hbox{/}_{T}$}}}
\def\DZ{D\O\ }

\def\ifb{fb$^{-1}$}

\def\pp{p\bar{p}}
\def\bb{b\bar{b}}
\def\cc{c\bar{c}}

\def\ttbar{$t\bar{t}$}

\def\lvbb{$\ell\nu\bb$}
\def\llbb{$\ell\ell\bb$}
\def\vvbb{$\nu\nu\bb$}

\def\ra{\rightarrow}

\def\wlv{W\ra\ell\nu}
\def\zll{Z\ra\ell\ell}
\def\zvv{Z\ra\nu\nu}

\def\wcs{W\ra c\bar{s}}
\def\zcc{Z\ra\cc}
\def\zbb{Z\ra\bb}

\def\gev{~Ge\kern -0.05em V\kern -0.1em /$c^2$}
\def\dgev{Ge\kern -0.05em V\kern -0.1em /$c^2$}
\newcommand{\GeV} {\ensuremath{\mathrm{Ge\kern -0.1em V}}}

\def\alpgen{{\sc alpgen}}
\def\pythia{{\sc pythia}}

\def\singletop{{\sc SingleTop}}
\def\mcfm{{\sc MCFM}}

\def\lumimin{7.5} 
\def\lumimax{9.5}

\def\vznlo{4.4}   
\def\vznloe{0.3}  

\def\wwnlo{11.3}  
\def\wwnloe{0.8}  
\def\wznlo{3.2}	  
\def\wznloe{0.2}  
\def\zznlo{1.2}   
\def\zznloe{0.1}  

\def\vzRF{4.47}

\def\vzRFstat{0.64}

\def\vzRFsystu{0.73}
\def\vzRFsystd{0.72}

\def\vzRFnsigma{4.6}

\def\vzresult{$\sigma(WW+WZ)=\vzRF\pm\vzRFstat$~(stat) $^{+\vzRFsystu}_{-\vzRFsystd}$~(syst)~pb}

\begin{document}





\rightline{FERMILAB-CONF-12-068-E}
\rightline{CDF Note 10802}
\rightline{\DZ Note 6311}
\vskip0.5in

\title{Combined CDF and \DZ measurement of $\bm{WZ}$ and $\bm{ZZ}$ production in final states with $\bm{b}$-tagged jets\\[2.5cm]}

\author{
The TEVNPH Working Group\footnote{The Tevatron
New-Phenomena and Higgs Working Group can be contacted at
TEVNPHWG@fnal.gov. More information can be found at http://tevnphwg.fnal.gov/.}
 }
\affiliation{\vskip0.3cm for the CDF and \DZ Collaborations\\ \vskip0.2cm
\today}





\begin{abstract}
\vskip0.3in  
  We present a combined measurement of the production cross section 
  of $VZ$ ($V=W$ or $Z$) events in final states containing charged
  leptons (electrons or muons) or neutrinos, and heavy flavor jets,
  using data collected by the CDF and D\O\ detectors at the Fermilab
  Tevatron Collider.  The analyzed samples of $p\bar{p}$ collisions at
$\sqrt{s}=1.96$ TeV correspond to integrated luminosities of \lumimin--\lumimax\ \ifb.  
  Assuming the ratio of the production cross sections $\sigma(WZ)$ 
  and $\sigma(ZZ)$ as predicted by the standard model, we measure 
  the sum of the $WZ$ and $ZZ$ cross sections to be \vzresult. This is consistent 
  with the standard model prediction and corresponds to a
  significance of \vzRFnsigma~standard deviations above the 
  background-only hypothesis.
\vspace*{4.0cm}
\end{abstract}

\maketitle
\centerline{\em Preliminary Results for the Moriond 2012 Conferences}

\newpage
\section{Introduction}
\label{intro}

\def\citeall{\cite{cdfWHl,cdfZHv,cdfZHl,dzWHl,dzZHv,dzZHl}}
\def\citealld0{\cite{dzWHl,dzZHv,dzZHl}}
\def\citeallcdf{\cite{cdfWHl,cdfZHv,cdfZHl}}
\def\citeallWHl{\cite{cdfWHl,dzWHl}}
\def\citeallZHv{\cite{cdfZHv,dzZHv}}
\def\citeallZHl{\cite{cdfZHl,dzZHl}}

Studies on the production of $VV$ ($V=W,Z$) boson pairs provide an 
important test of the electroweak sector of the standard model (SM).  
In $p\bar{p}$ collisions at $\sqrt{s}=1.96$ TeV, the next-to-leading
order (NLO) SM cross sections for these processes are
$\sigma(WW)=\wwnlo\pm\wwnloe$~pb, $\sigma(WZ)=\wznlo\pm\wznloe$~pb and
$\sigma(ZZ)=\zznlo\pm\zznloe$~pb~\cite{dibo}.  These cross sections
assume both $\gamma^{*}$ and $Z^{\circ}$ components in the neutral 
current exchange and corresponding production of dilepton final 
states in the region 75~$\leq m_{\ell^+\ell^-} \leq$~105~GeV/$c^2$.  
Measuring a significant departure in cross section or deviations in 
the predicted kinematic distributions would indicate the presence of 
anomalous gauge boson couplings~\cite{bib:anocoups} or new particles 
in extensions of the SM~\cite{bib:newphen}.  The $VV$ production in 
$\pp$ collisions at the Fermilab Tevatron Collider has been observed 
in fully leptonic decay modes~\cite{bib:leptonic} and in semi-leptonic 
decay modes~\cite{bib:hadronic}, where the combined $WW+WZ$ cross 
section was measured.  

Recently, the D\O\ experiment presented evidence for $WZ$ and $ZZ$ production 
in semileptonic decays with a $b$-tagged final state~\cite{dzDibosonCombo}.  
The $WZ$ and $ZZ$ production cross sections, as well as their sum,
were measured in final states where one of the $Z$ bosons decays into $\bb$ 
(although there is some signal contribution from $\wcs$, $\zcc$) and the other weak 
boson decays to charged leptons or neutrinos ($\wlv$, $\zvv$, or $\zll$, with $\ell=e,\mu$). 
In this note we report an improved measurement of the $WW+ZZ$ production cross section
in such final states based on the combination of the \DZ results from~\cite{dzDibosonCombo},
with a corresponding new set of CDF analyses~\cite{cdfDibosonCombo}.
This analysis is relevant as a proving ground for the combined Tevatron search for a low-mass 
Higgs boson produced in association with a weak boson and decaying into a $\bb$ pair
\cite{bib:higgs} since it shares the same selection criteria as well as analysis and combination techniques.

\section{Summary of Contributing Analyses}
\label{analyses}

This result is the combination of three CDF analyses~\citeallcdf~and 
three \DZ analyses~\citealld0~outlined in Table~\ref{tab:chans}.  These 
analyses utilize data corresponding to integrated luminosities ranging 
from \lumimin\ to \lumimax~\ifb, collected with the CDF~\cite{cdf} and
\DZ~\cite{dzero} detectors at the Fermilab Tevatron Collider, and they are 
organized into multiple sub-channels for each different configuration 
of final state particles.  To facilitate proper combination of signals, 
the analyses from a given experiment are constructed to use mutually 
exclusive event selections.

In the \lvbb~analyses~\citeallWHl, events containing an isolated 
electron or muon, and two or three jets are selected (exactly two 
jets in the case of the CDF analysis).  The presence of a neutrino 
from the $W$ decay is inferred from a large imbalance of transverse 
momentum ($\met$).  The \vvbb~analyses~\citeallZHv~select events 
containing large $\met$ and two or three jets (exactly two jets in the 
case of the \DZ analysis).  Finally, in the \llbb~analyses~\citeallZHl~ 
events are required to contain two electrons or two muons and at least 
two jets. In the case of the CDF  \llbb\ analysis, events with two or 
three jets are analyzed separately. In the  \DZ \lvbb\ and \llbb\ 
analyses as well as the CDF \llbb\ analysis, each 
lepton flavor of the $W/Z$ boson decay ($\ell=e,\mu$) is treated as an 
independent channel.  In the case of the CDF \lvbb\ analysis lepton 
types are separated into four different channels based on their purity 
and location within the detector.  To ensure that event samples used
for the different analyses do not overlap, the \lvbb\ analyses reject 
events in which a second isolated electron or muon is identified, and 
the \vvbb\ analyses reject events in which any isolated electrons or 
muons are identified.

To isolate the $\zbb$ decays, algorithms for identifying jets consistent 
with the decay of a heavy-flavor quark are applied to the jets in each 
event candidate ($b$-tagging).  All of the \DZ analyses, as well as 
the CDF \lvbb\ and \llbb\ analyses, use multivariate discriminants based 
on sets of kinematic variables sensitive to displaced decay vertices and 
tracks within jets with large transverse impact parameters relative to 
the hard-scatter vertices.  The \DZ algorithm is a boosted decision tree
discriminant which builds upon the previously utilized neural network 
$b$-tagging tool~\cite{bib:btagnn}, while the CDF algorithm~\cite{bib:HOBIT} 
is based on a neural network discriminant.  In both cases, a spectrum of 
increasingly stringent $b$-tagging operating points is constructed through 
the use of progressively higher requirements on the minimum
output of the $b$-tagging discriminant.  The \DZ analyses are separated 
into two groups: a double-tag (DT) group in which two of the jets are 
$b$-tagged with a loose tag requirement (\lvbb\ and \vvbb) or one loose 
and one tight tag requirement (\llbb); and an orthogonal single-tag (ST) 
group in which only one jet has a loose (\lvbb\ and \vvbb) or tight 
(\llbb) $b$-tag.  A typical per-jet $b$ efficiency and fake rate for the \DZ
loose (tight) $b$-tag selection is about 80\% (50\%) and 10\% (0.5\%), 
respectively.  The corresponding efficiency for jets from $c$-quarks 
is 45\% (12\%).  The \DZ \lvbb\ and \vvbb\ analyses also use the output 
of the $b$-tagging algorithm as an additional input to the discriminants 
used in the final signal extraction.  Candidate events in the CDF \lvbb\
and \llbb\ analyses are also separated into channels based on tight and 
loose tagging definitions.  Events with two tight tags (TT), one tight 
and one loose tag (TL), two loose tags (LL), and a single tight tag (Tx)
are used by both analyses.  The CDF \lvbb\ analysis also considers 
events with a single loose tag (Lx).   A typical per-jet efficiency and 
fake rate for the CDF loose (tight) neural network $b$-tag selection 
is about 70\% (45\%) and 7\% (0.6\%), respectively.  The CDF \vvbb\ 
analysis utilizes a tight $b$-tagging algorithm~\cite{bib:SecVtx} based on 
reconstruction of a displaced secondary vertex and a loose $b$-tagging 
algorithm~\cite{bib:JetProb} that assigns a likelihood for the tracks 
within a jet to have originated from a displaced vertex.  Based on the 
output of these algorithms events with two tight tags (SS) and those with 
one tight tag and one loose tag (SJ) are separated into independent analysis 
channels.  The signal in all of the double-tag samples is expected to be primarily composed 
of events with $\zbb$ decays, with smaller contributions from 
$\zcc$ and $\wcs$ decays.  In the single-tag samples, which are defined 
by less stringent requirements on the $b$-jet content of the event, the 
contributions from the three decay modes are comparable.  

\begin{table}[bp]
\caption{\label{tab:chans}List of analysis channels and their corresponding
integrated luminosities.  See Sect.~\ref{analyses} for details ($\ell=e, \mu$).}
\begin{ruledtabular}
\begin{tabular}{lccc}
\\
Experiment   &   Channel                               & Luminosity (\ifb) & Reference\\\hline
CDF          &   \lvbb,~ TT/TL/Tx/LL/Lx, 2 jets        & 9.5               & \cite{cdfWHl}\\
CDF          &   \vvbb,~ SS/SJ, 2/3 jets               & 9.5               & \cite{cdfZHv}\\
CDF          &   \llbb,~ TT/TL/Tx/LL, 2/3 jets         & 9.5               & \cite{cdfZHl}\\
\DZ           &   \lvbb,~ ST/DT, 2/3 jets               & 7.5               & \cite{dzWHl}\\
\DZ           &   \vvbb,~ ST/DT, 2 jets                 & 8.4               & \cite{dzZHv}\\
\DZ           &   \llbb,~ ST/DT, $\geq$ 2 jets          & 7.5               & \cite{dzZHl}\\
\end{tabular}
\end{ruledtabular}
\end{table}

The primary background is from $W/Z$+jets, which is modeled with
\alpgen~\cite{alpgen} by both CDF and D\O.  The backgrounds from 
multijet production are measured from control samples in the data.  
At D\O\ the other backgrounds are generated with \alpgen~and
\singletop~\cite{singletop}, with \pythia~\cite{pythia} providing
parton-showering and hadronization.  At CDF most backgrounds from
other SM processes are modeled using \pythia\ Monte Carlo samples.  Background rates 
are normalized either to next-to-leading order (NLO) or higher-order
theory calculations or to data control samples.  The D\O\ \llbb\ and
both experiment's \lvbb\ analyses normalize $W/Z$+jets backgrounds 
to data, whereas the the CDF \llbb\ and both experiment's \vvbb\ 
analyses normalize them to the predictions from \alpgen.  The
fraction of the $W/Z$+jets in which the jets arise from heavy quarks
($b$ or $c$) is obtained from NLO calculations using \mcfm~\cite{mcfm}
at D\O\ while at CDF the prediction from \alpgen\ is corrected based
on a data control region.  The background from \ttbar\ events is
normalized to the approximate NNLO cross section \cite{ttbar_xsec}.
The $s$-channel and $t$-channel cross sections for the production of
single-top quarks are from approximate NNLO+NNLL calculations
\cite{schan_top_xsec} and approximate NNNLO+NLL calculations
\cite{tchan_top_xsec}, respectively.  The background from $WW$ events
is normalized to NLO calculations from \mcfm~\cite{dibo}.
All Monte Carlo samples
are passed through detailed {\sc geant}-based simulations~\cite{geant} of the CDF and D0 detectors.

The \DZ analyses use multivariate discriminants (MVA) based on decision 
trees as the final variables for extracting the $VZ$ signal from the backgrounds. 
These decision trees are trained to discriminate the $VZ$ signal from the backgrounds using
the same set of discriminant variables as in the corresponding Higgs analyses. 
The CDF analyses follow the same strategy, using neural network-based discriminants
instead for signal-to-background discrimination.

\section{Systematic Uncertainties}

Systematic uncertainties differ between experiments and analyses, and 
they affect the normalizations and the differential distributions
(shapes) of the predicted signal and background 
templates in correlated ways.  The combined result incorporates the 
sensitivity of predictions to values of nuisance parameters and takes 
into account correlations in these parameters both within each individual 
experiment and between experiments.  The largest uncertainty contributions 
and their correlations between and within the two experiments are discussed
here.  Further details on the individual analyses are available
in Refs.~\citeall.  

\subsubsection{Correlated Systematics between CDF and \DZ}

The uncertainties on measurements of the integrated luminosities are 
5.9\% (CDF) and 6.1\% (\DZ).  Of these values, 4\% arises from the
uncertainty on the inelastic $\pp$~scattering cross section, which 
is correlated between CDF and D\O.  CDF and \DZ also share the assumed
values and uncertainties on the cross sections for $WW$ production 
and top-quark production processes (\ttbar~and single top).

In most analyses determination of the multijet (``QCD'')
background involves data control samples, and the methods used 
differ between CDF and D\O, and even between analyses within the 
collaborations.  Therefore, there is no assumed correlation in 
the predicted rates of this background between analysis channels. 
Likewise, calibrations of quantities such as the fake lepton rate, 
$b$-tag efficiencies, and mistag rates are performed by each 
collaboration using independent data samples and different methods, and are 
treated as uncorrelated. Similarly, different techniques are used 
to estimate background rates for $W/Z$+heavy flavor backgrounds 
and the associated uncertainties are taken as uncorrelated.

\subsubsection{Correlated Systematic Uncertainties for CDF}
The dominant systematic uncertainties for the CDF analyses are shown
in Appendix Tables~\ref{tab:cdfsystwh1} and~\ref{tab:cdfsystwh2} 
for the \lvbb\ channels, in Table~\ref{tab:cdfsystzhvv} for the \vvbb\
channels, and in Tables~\ref{tab:cdfllbb1} and~\ref{tab:cdfllbb2} 
for the \llbb\ channels.  Each source induces a
correlated uncertainty across all of CDF's channels' signal and background
contributions which are sensitive to that source.  The
largest uncertainties on signal arise from measured $b$-tagging
efficiencies, jet energy scale, and other Monte Carlo modeling.  Shape
dependencies of templates on jet energy scale, $b$-tagging, and gluon
radiation (``ISR'' and ``FSR'') are taken into account for some
analyses (see tables).  Uncertainties on background event rates vary
significantly for the different processes.  The backgrounds with the
largest systematic rate uncertainties are in general quite small. Such
uncertainties are constrained through fits to the nuisance parameters 
and do not affect the result significantly.  Since normalizations for 
the $W/Z$+heavy flavor backgrounds are obtained from data in the \lvbb\
and \vvbb\ analyses, the corresponding rate uncertainties associated 
with each analysis are treated as uncorrelated even within CDF.

\subsubsection{Correlated Systematic Uncertainties for \DZ }

The \vvbb\ and \lvbb\ analyses carry an uncertainty on
the integrated luminosity of 6.1\%~\cite{lumi}, while the overall
normalization of the \llbb\ analysis is determined from the NNLO
$Z/\gamma^*$ cross section \cite{dyxsec} in data events near the peak
of $\zll$ decays.  The uncertainty from the identification and energy 
measurement of jets is $\sim$7\%.  The uncertainty arising from the
$b$-tagging rate ranges from 1 to 10\%.  All analyses include
uncertainties associated with lepton measurement and acceptances,
which range from 1 to 9\% depending on the final state.  The largest
contribution for all analyses is the theoretical uncertainty on the
background cross sections at 7-20\% depending on the analysis channel
and specific background.  The uncertainty on the expected multijet
background is dominated by the statistics of the data sample from
which it is estimated.  
Further details on the systematic uncertainties are given in
Tables~\ref{tab:d0systwh}-\ref{tab:d0llbb1}.  All systematic
uncertainties originating from a common source are taken
to be 100\% correlated, as detailed in Table \ref{tab:corr}.

\section{Measurement of the $\bm{WZ+ZZ}$ Cross Section}

The total $VZ$ cross section is determined from a maximum likelihood
fit of the MVA distributions for the background and signal samples
from the contributing analyses to the data.  The cross section for the
signal ($WZ+ZZ$) is a free parameter in the fit, but the ratio of the
$WZ$ and $ZZ$ cross sections is fixed to the SM prediction.  Events
from $WW$ production are considered as a background. The fit is
performed simultaneously on the distributions in all sub-channels.  As
a consistency check, we also determine the Bayesian posterior
probability by integrating over the nuisance parameters.  Here we
report only the results from the maximum likelihood fit, but the
results from the Bayesian method are consistent.

\begin{figure*}[tbp]
\begin{centering}
\includegraphics[height=0.2\textheight]{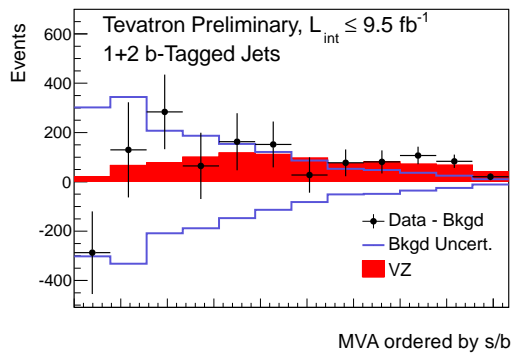}
\end{centering} 
\caption{\label{fig:rfsub} 
Comparison of the measured $VZ$ signal (filled histogram) to
background-subtracted data (points) after the maximum likelihood fit.
The distribution is a combination of all final discriminants where the
bins are ordered and merged according to their expected signal to background
ratio ($s/b$). The $x$-axis has arbitrary units.  Also
shown is the $\pm$1 standard deviation uncertainty on the fitted
background that was subtracted.
}
\end{figure*}

The combined fit for the total $VZ$ cross section distributions yields
\vzresult.  This measurement is consistent with the NLO SM prediction
of $\sigma(WW+WZ)=\vznlo\pm\vznloe$~pb~\cite{dibo}, as well as
with the individual measurements from \DZ\cite{dzDibosonCombo},
$\sigma(WW+WZ)=5.0 \pm 1.6$~pb, and 
from CDF~\cite{cdfDibosonCombo},
$\sigma(WW+WZ)=4.1 ^{+1.4}_{-1.3}$~pb.
Based on the measured central value for the $VZ$ cross section
and its uncertainties, the observed significance is estimated to be
\vzRFnsigma~standard deviations (s.d.), while the expected
significance is $\sim 4.8$ s.d.

To visualize the sensitivity of the combined analysis, we calculate
the expected signal over background ($s/b$) in each bin of the MVA distributions
from the contributing analyses.  Bins with similar $s/b$ are then
combined to produce a single distribution, shown in Fig.~\ref{fig:rfsub}.
The binning was chosen to keep the background fluctuations roughly of 
the same size as in the dijet mass distributions.
Figure~\ref{fig:mjj} shows the distributions of the invariant
mass of the dijet system, summed over all channels from CDF and
D\O, after adjusting the signal and background
predictions according to the results of the fit.  Figure
\ref{fig:mjj_sub} shows the background subtracted dijet mass
distributions after the fit, demonstrating the presence of a hadronic
resonance in the data consistent with the SM expectation, both
in shape and normalization.

\begin{figure*}[tp]
\begin{centering}
\begin{tabular}{ccc}
\includegraphics[width=2.4in]{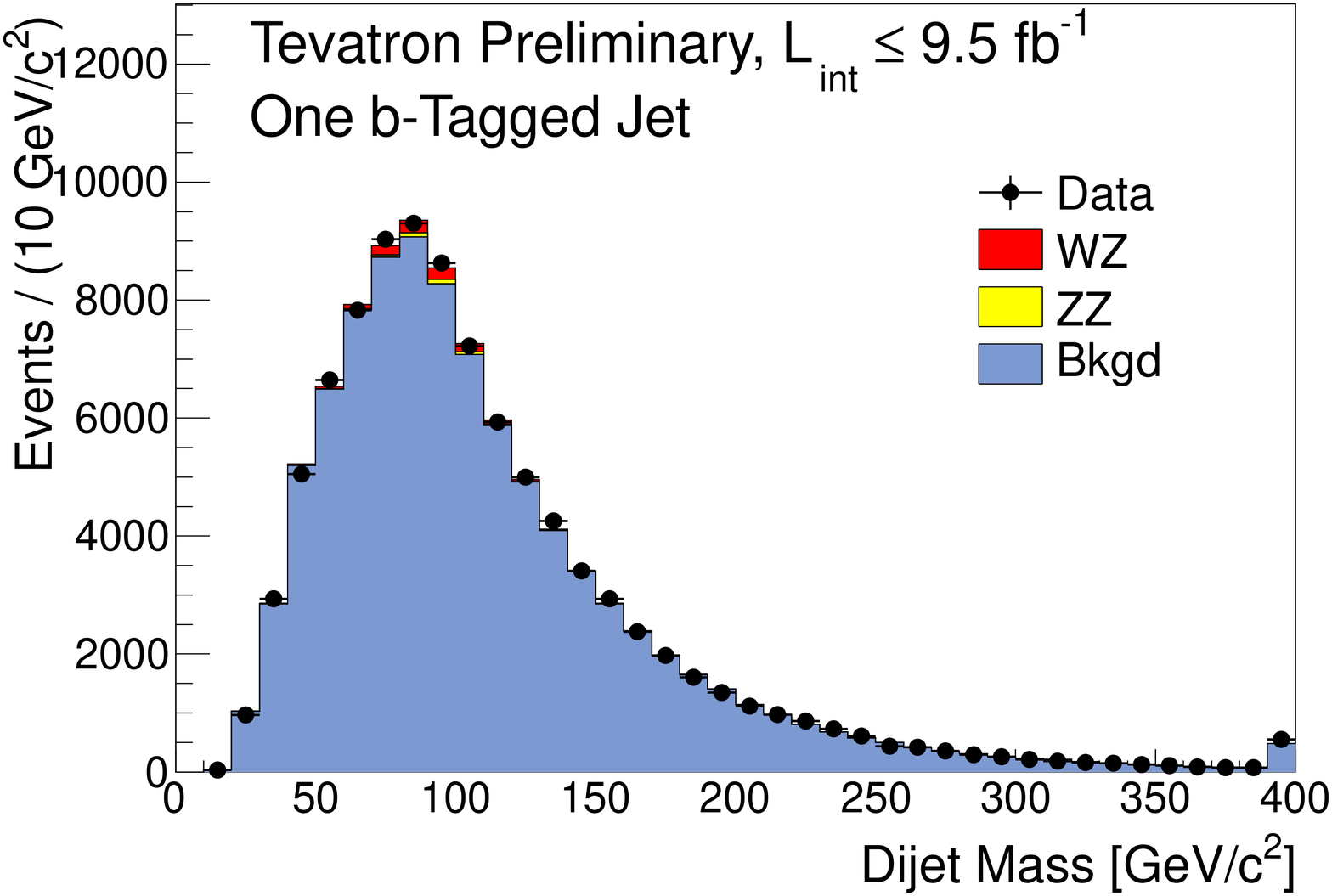}  &
\includegraphics[width=2.4in]{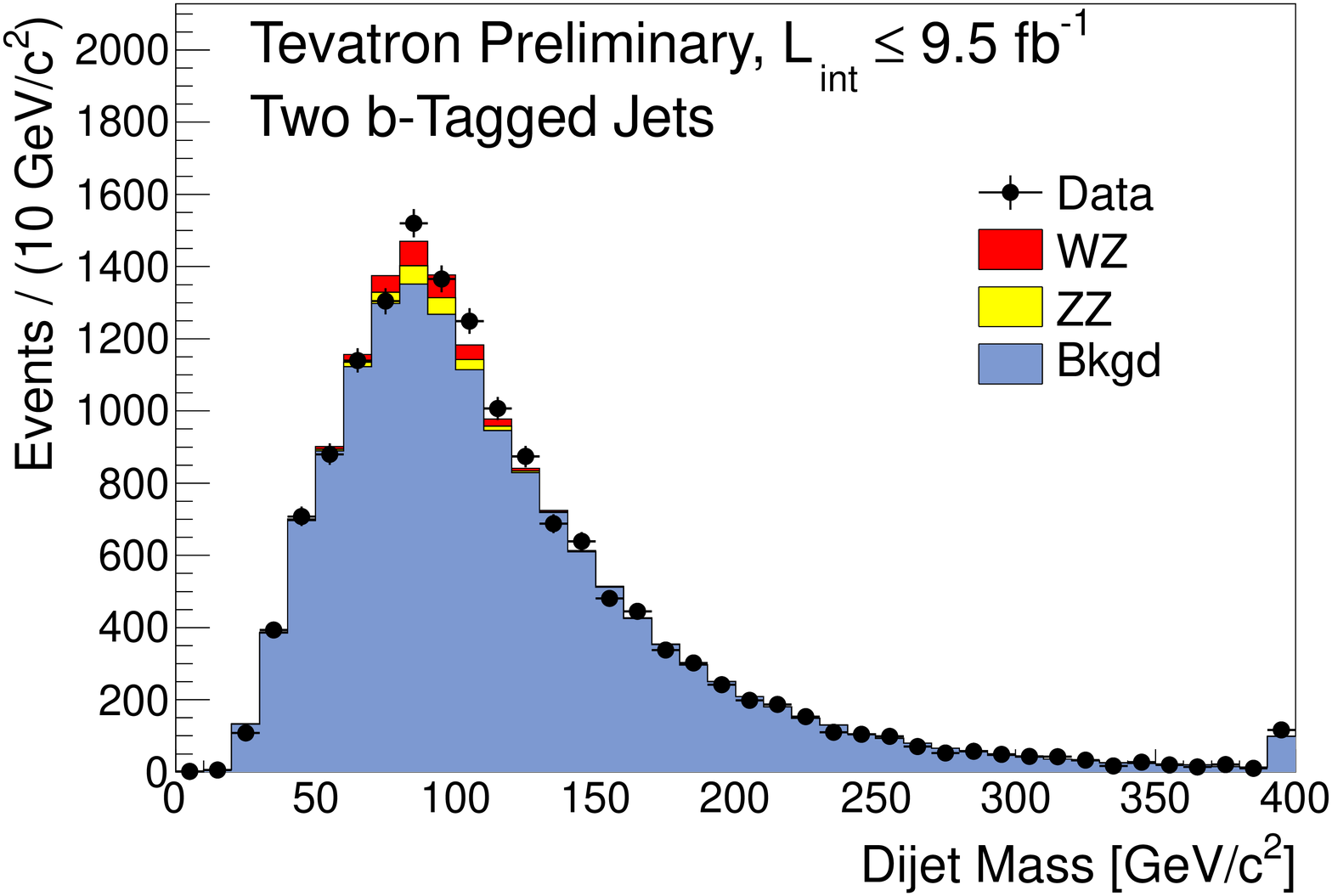} & 
\includegraphics[width=2.4in]{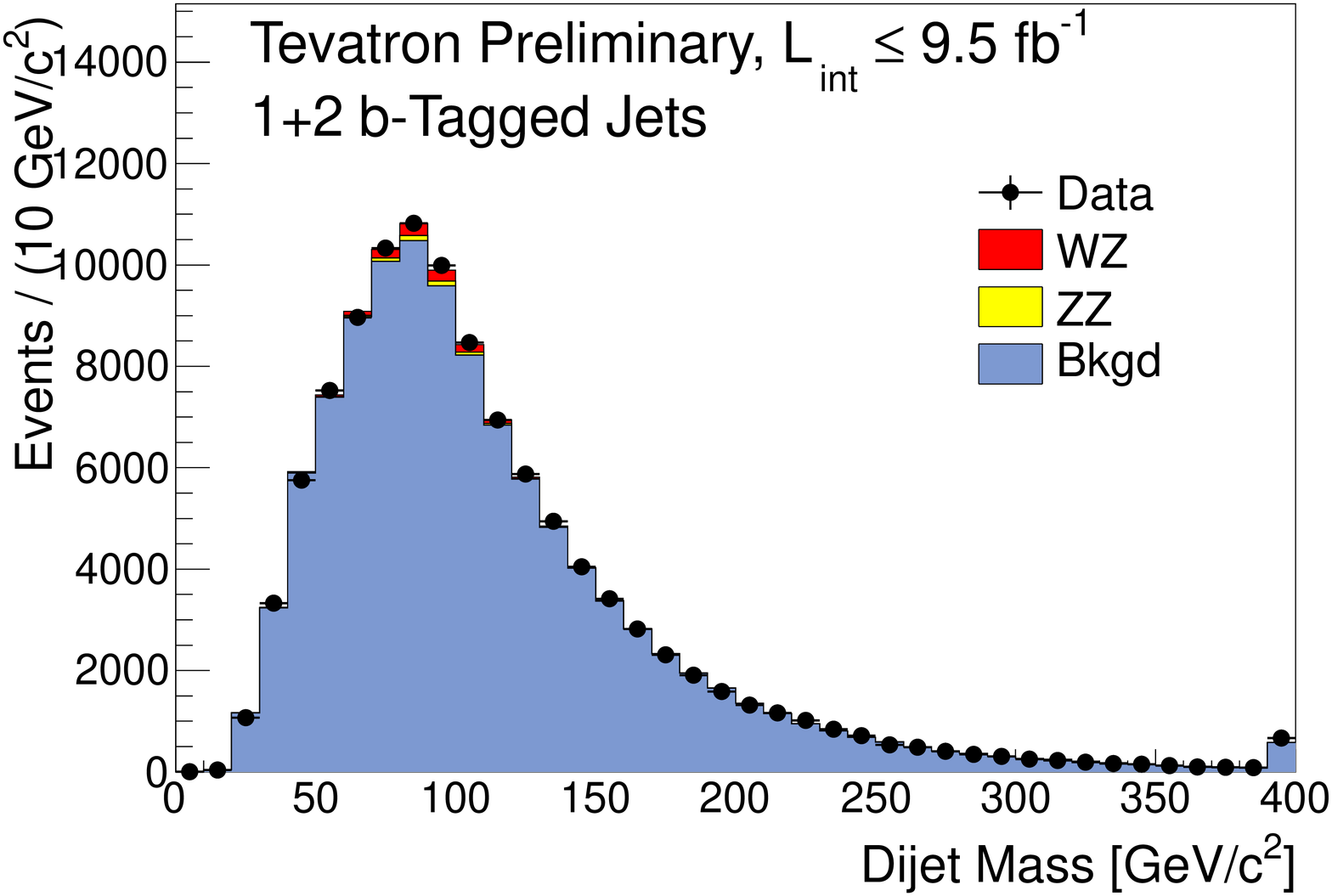} \\
{\bf (a)} & {\bf (b)}  & {\bf (c)}
\end{tabular} 
\end{centering} 
\caption{\label{fig:mjj} Comparison of the fitted signal+background to
	data in the dijet mass distribution (summed over all channels) 
        for the (a) ST, and (b) DT sub-channels; and (c) the sum of the
        ST and DT sub-channels.  Events with a dijet mass greater
        than 400 GeV are included in the last bin of the distribution.}
\end{figure*}
\begin{figure*}[tp]
\begin{centering}
\begin{tabular}{ccc}
\includegraphics[width=2.4in]{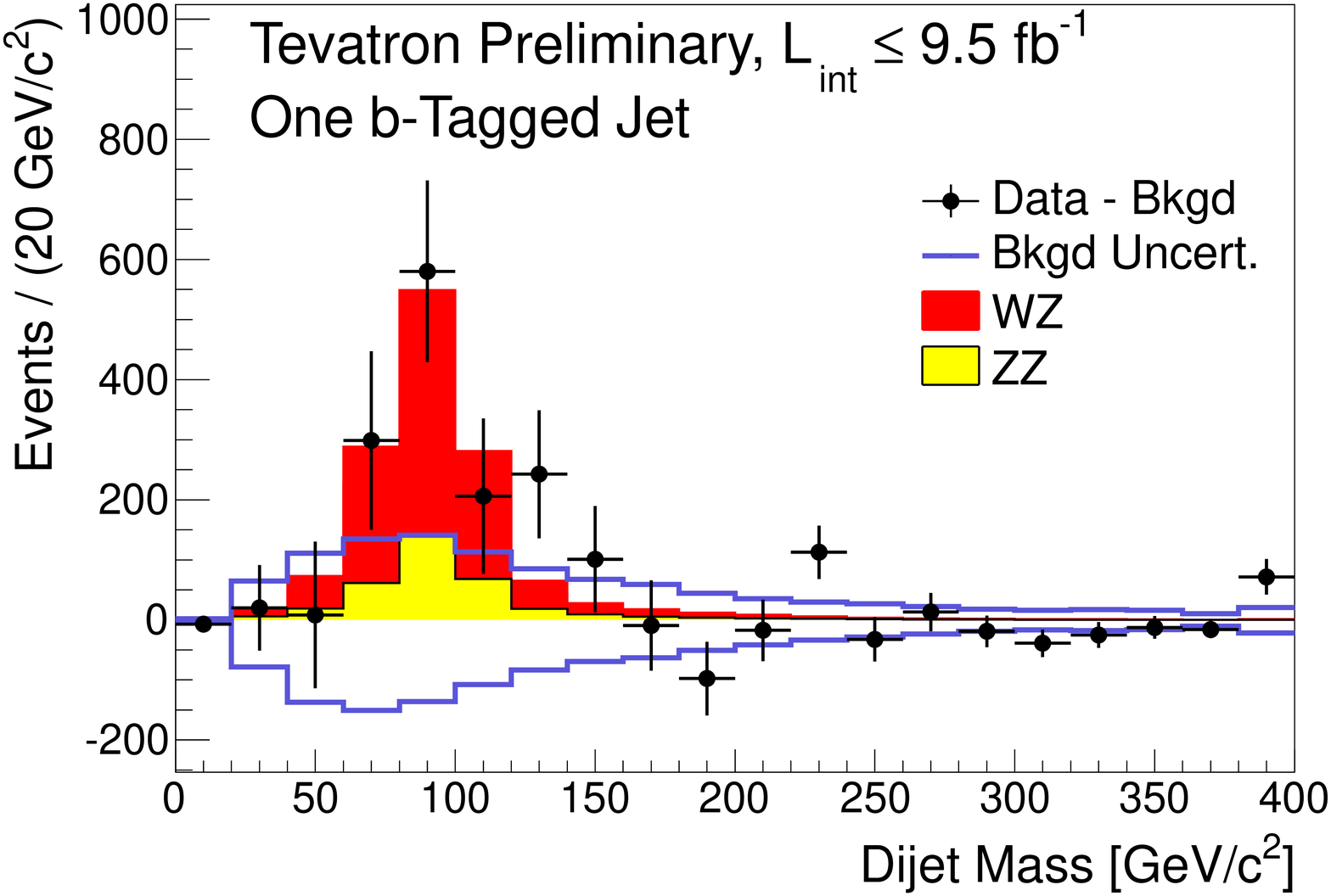}  &
\includegraphics[width=2.4in]{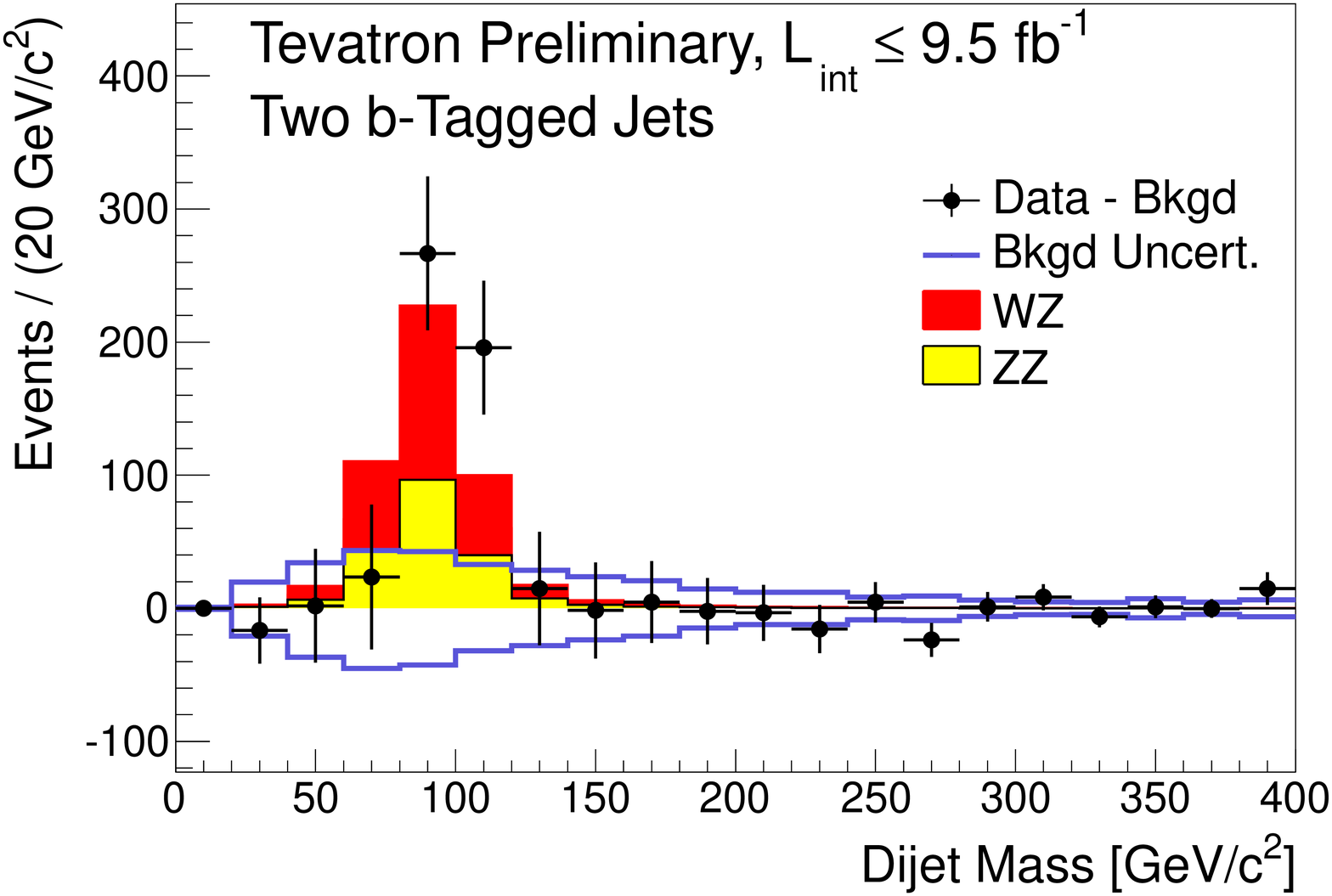} & 
\includegraphics[width=2.4in]{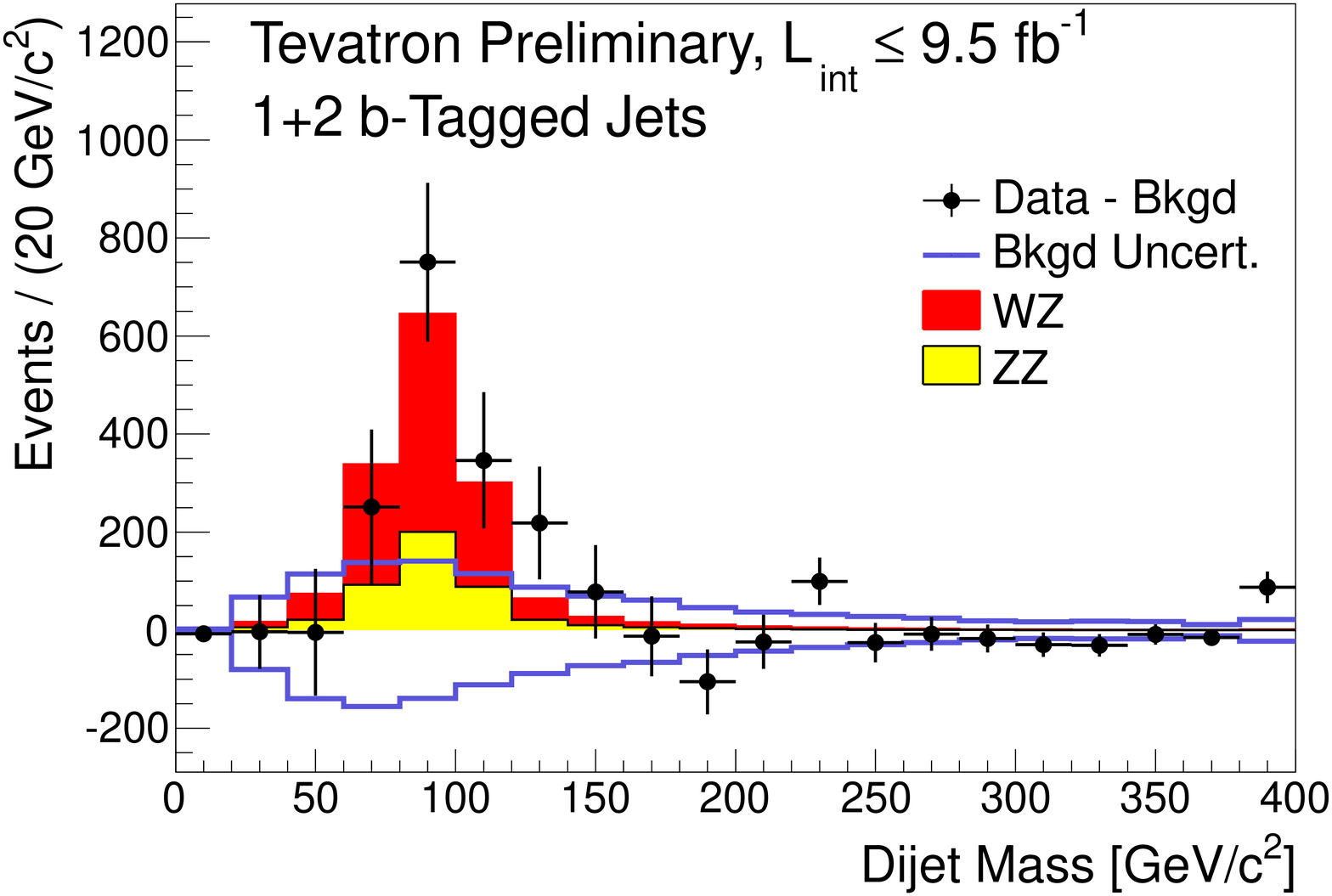} \\
{\bf (a)} & {\bf (b)}  & {\bf (c)}
\end{tabular} 
\end{centering} 
\caption{\label{fig:mjj_sub} Comparison of the measured $WZ$ and
	$ZZ$ signals (filled histograms) to background-subtracted data
(points) in the dijet mass distribution (summed over all channels) 
        for the (a) ST, and (b) DT sub-channels; and (c) the sum of the
        ST and DT sub-channels. Also shown is the $\pm$1 standard deviation
        uncertainty on the fitted background.  Events with a dijet mass greater
        than 400 GeV are included in the last bin of the distribution.}
\end{figure*}

\section{Summary}

In summary, we combine analyses in the \lvbb, \vvbb, and
\llbb\ ($\ell=e,~\mu$) final states from the CDF and D\O\ experiments
to observe, with a significance of \vzRFnsigma~s.d., the production of
$VZ$ ($V=W$ or $Z$) events.  The analyzed samples correspond to
\lumimin\ to \lumimax\ \ifb~of $\pp$~collisions at $\sqrt{s}=1.96$
TeV.  We measure the total cross section for $VZ$ production to be
\vzresult. This result demonstrates the ability of the Tevatron
experiments to measure a SM production process with cross section 
of the same order magnitude as that expected for Higgs production 
from the same set of background-dominated final states containing 
two heavy-flavor jets used in our low mass Higgs searches.

\newpage
\begin{acknowledgments}
\input{acknowledgement}
\end{acknowledgments}

\newpage
\appendix

\section{Additional Material}

\input{cdf-lvbb-sys}

\input{cdf-vvbb-sys}

\input{cdf-llbb-sys}

\input{d0-lvbb-sys}

\input{d0-vvbb-sys}

\input{d0-llbb-sys}

\begin{table}[htpb]
  \caption{\label{tab:corr}The correlation matrix for the D0 analysis
    channels.  Uncertainties marked with an $\times$ are considered
    100\% correlated across the affected channels.  Otherwise
    the uncertainties are not considered correlated, or do not
    apply to the specific channel.  The systematic uncertainties
    on the background cross section ($\sigma$) and the normalization
    are each subdivided according to the different background
    processes in each analysis. }
\begin{ruledtabular}
\begin{tabular}{lcccc}\\
Source                     & \lvbb\      & \vvbb\    & \llbb\      \\ \hline
Luminosity                 & $\times$   & $\times$   &              \\
Normalization	           &            &	     &              \\
Jet Energy Scale           & $\times$   & $\times$   & $\times$     \\
Jet ID                     & $\times$   & $\times$   & $\times$     \\
Electron ID/Trigger        & $\times$   & $\times$   & $\times$     \\
Muon ID/Trigger            & $\times$   & $\times$   & $\times$     \\
$b$-Jet Tagging            & $\times$   & $\times$   & $\times$     \\
Background $\sigma$        & $\times$   & $\times$   & $\times$     \\
Background Modeling        &            &            &              \\
Multijet Background        &            &            &              \\
Signal $\sigma$            & $\times$   & $\times$   &  $\times$    \\
\hline
\\
\end{tabular}
\end{ruledtabular}
\end{table}

\end{document}

%% file: acknowledgement.tex
%
We thank the Fermilab staff and the technical staffs of the
participating institutions for their contributions, and we
acknowledge support from the
DOE and NSF (USA);
CONICET and UBACyT (Argentina);
ARC (Australia);
CNPq, FAPERJ, FAPESP and FUNDUNESP (Brazil);
CRC Program and NSERC (Canada);
CAS, CNSF, and NSC (China);
Colciencias (Colombia);
MSMT and GACR (Czech Republic);
Academy of Finland (Finland);
CEA and CNRS/IN2P3 (France);
BMBF and DFG (Germany);
INFN (Italy);
DAE and DST (India);
SFI (Ireland);
Ministry of Education, Culture, Sports, Science and Technology (Japan);
KRF, KOSEF and World Class University Program (Korea);
CONACyT (Mexico);
FOM (The Netherlands);
FASI, Rosatom and RFBR (Russia);
Slovak R\&D Agency (Slovakia);
Ministerio de Ciencia e Innovaci\'{o}n, and Programa Consolider-Ingenio 2010 (Spain);
The Swedish Research Council (Sweden);
Swiss National Science Foundation (Switzerland);
STFC and the Royal Society (United Kingdom);
and
the A.P. Sloan Foundation (USA).

%% file: cdf-lvbb-sys.tex
\begin{table}[h]
\begin{center}
  \caption{\label{tab:cdfsystwh1} Systematic uncertainties for the CDF
    $\ell\nu b{\bar{b}}$ single tight tag (Tx) and single loose tag (Lx) 
    channels.  Systematic uncertainties are listed by name; see the 
    original references for a detailed explanation of their meaning and 
    on how they are derived.  Uncertainties are relative, in percent on 
    the event yield.  Shape uncertainties are labeled with an ``(S)''.}
\vskip 0.2cm
{\centerline{CDF $\ell\nu b\bar{b}$ single tight tag (Tx) channels relative uncertainties (\%)}}
\vskip 0.099cm
\begin{ruledtabular}
\begin{tabular}{lcccccc}\\
Contribution              & $W$+HF & Mistags & Top & Diboson & Non-$W$ & $WH$  \\ \hline
Luminosity ($\sigma_{\mathrm{inel}}(p{\bar{p}})$)
                          & 3.8      & 0       & 3.8 & 3.8     & 0       &    3.8   \\
Luminosity Monitor        & 4.4      & 0       & 4.4 & 4.4     & 0       &    4.4   \\
Lepton ID                 & 2.0-4.5      & 0       & 2.0-4.5   & 2.0-4.5       & 0       &    2.0-4.5   \\
Jet Energy Scale          & 3.2-6.9(S)      & 0.9-1.8(S)       & 0.8-9.7(S)   & 3.6-13.2(S)       & 0       &    3.0-5.0(S)   \\
Mistag Rate (tight)               & 0      & 19    & 0   & 0       & 0       &    0   \\
Mistag Rate (loose)               & 0      & 0    & 0   & 0       & 0       &    0   \\
$B$-Tag Efficiency (tight)         & 0      & 0       & 3.9 & 3.9     & 0       &    3.9   \\
$B$-Tag Efficiency (loose)        & 0      & 0       & 0 & 0     & 0       &    0   \\
$t{\bar{t}}$ Cross Section  & 0    & 0       & 10  & 0       & 0       &    0   \\
Diboson Rate                & 0      & 0       & 0   & 6.0    & 0       &    0   \\
Signal Cross Section        & 0      & 0       & 0   & 0       & 0       &    5 \\
HF Fraction in W+jets       &    30  & 0       & 0   & 0       & 0       &    0   \\
ISR+FSR+PDF                 & 0      & 0       & 0   & 0       & 0       &    3.8-6.8 \\
$Q^2$                       &  3.2-6.9(S)           &    0.9-1.8(S)     &    0       &   0            & 0        &    0        \\
QCD Rate                    & 0      & 0       & 0   & 0       & 40      &    0   \\
\end{tabular}
\end{ruledtabular}
\vskip 0.5cm
{\centerline{CDF $\ell\nu b\bar{b}$ single loose tag (Lx) channels relative uncertainties (\%)}}
\vskip 0.099cm
\begin{ruledtabular}
\begin{tabular}{lcccccc}\\
Contribution              & $W$+HF & Mistags & Top & Diboson & Non-$W$ & $WH$  \\ \hline
Luminosity ($\sigma_{\mathrm{inel}}(p{\bar{p}})$)
                          & 3.8      & 0       & 3.8 & 3.8     & 0       &    3.8   \\
Luminosity Monitor        & 4.4      & 0       & 4.4 & 4.4     & 0       &    4.4   \\
Lepton ID                 & 2      & 0       & 2   & 2       & 0       &    2   \\
Jet Energy Scale          & 2.2-6.0(S)      & 0.9-1.8(S)       & 1.6-8.6(S)   & 4.6-9.6(S)       & 0       &    3.1-4.8(S)   \\
Mistag Rate (tight)               & 0      & 0    & 0   & 0       & 0       &    0   \\
Mistag Rate (loose)               & 0      & 10    & 0   & 0       & 0       &    0   \\
$B$-Tag Efficiency (tight)         & 0      & 0       & 0 & 0     & 0       &    0   \\
$B$-Tag Efficiency (loose)        & 0      & 0       & 3.2 & 3.2     & 0       &    3.2   \\
$t{\bar{t}}$ Cross Section  & 0    & 0       & 10  & 0       & 0       &    0   \\
Diboson Rate              & 0      & 0       & 0   & 6.0    & 0       &    0   \\
Signal Cross Section      & 0      & 0       & 0   & 0       & 0       &    10 \\
HF Fraction in W+jets     &    30  & 0       & 0   & 0       & 0       &    0   \\
ISR+FSR+PDF               & 0      & 0       & 0   & 0       & 0       &    2.4-4.9 \\
QCD Rate                  & 2.1-6.0(S)      & 0.9-1.8(S)       & 0   & 0       & 40      &    0   \\
\end{tabular}
\end{ruledtabular}

\end{center}
\end{table}

\begin{table}[h]
\begin{center}
  \caption{\label{tab:cdfsystwh2} Systematic uncertainties for the CDF
    $\ell\nu b{\bar{b}}$ double tight tag (TT), one tight tag and one 
    loose tag (TL) and double loose tag (LL) channels.  Systematic 
    uncertainties are listed by name; see the original references for 
    a detailed explanation of their meaning and on how they are derived.  
    Uncertainties are relative, in percent on the event yield.  Shape 
    uncertainties are labeled with an ``(S)''.}
\vskip 0.2cm
{\centerline{CDF $\ell\nu b\bar{b}$ double tight tag (TT) channels relative uncertainties (\%)}}
\vskip 0.099cm
\begin{ruledtabular}
\begin{tabular}{lcccccc}\\
Contribution              & $W$+HF & Mistags & Top & Diboson & Non-$W$ & $WH$  \\ \hline
Luminosity ($\sigma_{\mathrm{inel}}(p{\bar{p}})$)
                          & 3.8      & 0       & 3.8 & 3.8     & 0       &    3.8   \\
Luminosity Monitor        & 4.4      & 0       & 4.4 & 4.4     & 0       &    4.4   \\
Lepton ID                 & 2.0-4.5      & 0       & 2.0-4.5   & 2.0-4.5       & 0       &    2.0-4.5   \\
Jet Energy Scale          & 4.0-16.6(S)      & 0.9-3.3(S)       & 0.9-10.4(S)   & 4.7-19.7(S)       & 0       &    2.3-13.6(S)   \\
Mistag Rate (tight)               & 0      & 40     & 0   & 0       & 0       &    0   \\
Mistag Rate (loose)               & 0      & 0     & 0   & 0       & 0       &    0   \\
$B$-Tag Efficiency (tight)         & 0      & 0       & 7.8 & 7.8     & 0       &    7.8   \\
$B$-Tag Efficiency (loose)        & 0      & 0       & 0 & 0     & 0       &    0   \\
$t{\bar{t}}$ Cross Section  & 0    & 0       & 10  & 0       & 0       &    0   \\
Diboson Rate              & 0      & 0       & 0   & 6.0   & 0       &    0   \\
Signal Cross Section      & 0      & 0       & 0   & 0       & 0       &    5 \\
HF Fraction in W+jets     &    30  & 0       & 0   & 0       & 0       &    0   \\
ISR+FSR+PDF               & 0      & 0       & 0   & 0       & 0       &    6.4-12.6 \\
$Q^2$                     & 4.0-8.8(S)          & 0.9-1.8(S)        & 0    &   0      &   0      &  0   \\
QCD Rate                  & 0      & 0       & 0   & 0       & 40      &    0   \\
\end{tabular}
\end{ruledtabular}
\vskip 0.5cm
{\centerline{CDF $\ell\nu b\bar{b}$ one tight and one loose tag (TL) channels relative uncertainties (\%)}}
\vskip 0.099cm
\begin{ruledtabular}
\begin{tabular}{lcccccc}\\
Contribution              & $W$+HF & Mistags & Top & Diboson & Non-$W$ & $WH$  \\ \hline
Luminosity ($\sigma_{\mathrm{inel}}(p{\bar{p}})$)
                          & 3.8      & 0       & 3.8 & 3.8     & 0       &    3.8   \\
Luminosity Monitor        & 4.4      & 0       & 4.4 & 4.4     & 0       &    4.4   \\
Lepton ID                 & 2.0-4.5      & 0       & 2.0-4.5   & 2.0-4.5       & 0       &    2.0-4.5   \\
Jet Energy Scale          & 3.9-12.4(S)      &  0.9-3.3(S)      & 1.4-11.5(S)   & 5.0-16.0(S)       &        &    2.5-16.1(S)   \\
Mistag Rate (tight)               & 0      & 19     & 0   & 0       & 0       &    0   \\
Mistag Rate (loose)               & 0      & 10     & 0   & 0       & 0       &    0   \\
$B$-Tag Efficiency (tight)         & 0      & 0       & 3.9 & 3.9     & 0       &    3.9   \\
$B$-Tag Efficiency (loose)        & 0      & 0       & 3.2 & 3.2     & 0       &    3.2   \\
$t{\bar{t}}$ Cross Section  & 0    & 0       & 10  & 0       & 0       &    0   \\
Diboson Rate              & 0      & 0       & 0   & 6.0    & 0       &    0   \\
Signal Cross Section      & 0      & 0       & 0   & 0       & 0       &    5 \\
HF Fraction in W+jets     &    30  & 0       & 0   & 0       & 0       &    0   \\
ISR+FSR+PDF               & 0      & 0       & 0   & 0       & 0       &    3.3-10.3 \\
$Q^2$                     & 3.9-7.7(S)             &  0.9-1.9(S)       &     0       &   0             & 0        & 0 \\
QCD Rate                  & 0      & 0       & 0   & 0       & 40      &    0   \\
\end{tabular}
\end{ruledtabular}
\vskip 0.5cm
{\centerline{CDF $\ell\nu b\bar{b}$ one tight and one loose tag (TL) channels relative uncertainties (\%)}}
\vskip 0.099cm
\begin{ruledtabular}
\begin{tabular}{lcccccc}\\
Contribution              & $W$+HF & Mistags & Top & Diboson & Non-$W$ & $WH$  \\ \hline
Luminosity ($\sigma_{\mathrm{inel}}(p{\bar{p}})$)
                          & 3.8      & 0       & 3.8 & 3.8     & 0       &    3.8   \\
Luminosity Monitor        & 4.4      & 0       & 4.4 & 4.4     & 0       &    4.4   \\
Lepton ID                 & 2      & 0       & 2   & 2       & 0       &    2   \\
Jet Energy Scale          & 3.6-6.9(S)      & 0.9-1.8(S)       & 1.7-7.9(S)   & 1.2-8.5       & 0       &    2.7-5.4(S)   \\
Mistag Rate (tight)               & 0      & 0     & 0   & 0       & 0       &    0   \\
Mistag Rate (loose)               & 0      & 20     & 0   & 0       & 0       &    0   \\
$B$-Tag Efficiency (tight)         & 0      & 0       & 0 & 0     & 0       &    0   \\
$B$-Tag Efficiency (loose)        & 0      & 0       & 6.3 & 6.3     & 0       &    6.3   \\
$t{\bar{t}}$ Cross Section  & 0    & 0       & 10  & 0       & 0       &    0   \\
Diboson Rate              & 0      & 0       & 0   & 6.0    & 0       &    0   \\
Signal Cross Section      & 0      & 0       & 0   & 0       & 0       &    10 \\
HF Fraction in W+jets     &    30  & 0       & 0   & 0       & 0       &    0   \\
ISR+FSR+PDF               & 0      & 0       & 0   & 0       & 0       &    2.0-13.6 \\
QCD Rate                  & 3.6-6.9(S)      & 0.9-1.8(S)       & 0   & 0       & 40      &    0   \\
\end{tabular}
\end{ruledtabular}

\end{center}
\end{table}

%% file: cdf-vvbb-sys.tex
\begin{table}[h]
  \caption{\label{tab:cdfsystzhvv} Systematic uncertainties for the CDF
    $\nu\nu b{\bar{b}}$ tight double tag (SS) and loose double tag (SJ) channels. 
    Systematic uncertainties are listed by name; see the original references 
    for a detailed explanation of their meaning and on how they are derived. 
    Uncertainties are relative, in percent on the event yield.  Shape 
    uncertainties are labeled with an ``(S)''.}
\vskip 0.2cm
{\centerline{CDF $\nu\nu b\bar{b}$ tight double tag (SS) channel relative uncertainties (\%)}}
\vskip 0.099cm
\begin{ruledtabular}
      \begin{tabular}{lccccccccc}\\
        Contribution & ZH & WH & Multijet & Mistags & Top Pair & S. Top  & Diboson  & W + HF  & Z + HF \\\hline
        Luminosity       & 3.8 & 3.8 &     &  & 3.8 & 3.8 & 3.8     & 3.8     & 3.8     \\
        Lumi Monitor      & 4.4 & 4.4 &     &  & 4.4 & 4.4 & 4.4     & 4.4     & 4.4     \\
        Tagging SF        & 10.4& 10.4&      & & 10.4& 10.4& 10.4    & 10.4    & 10.4    \\
      Trigger Eff. (S)& 0.9 & 1.4 & 0.9 & & 0.9 & 1.6 & 2.0     & 1.8     & 1.2     \\
        Lepton Veto       & 2.0 & 2.0 &      & & 2.0 & 2.0 &2.0      & 2.0     & 2.0     \\
        PDF Acceptance    & 3.0 & 3.0 &    &   & 3.0 & 3.0 &3.0      & 3.0     & 3.0     \\
        JES (S)       & $^{+1.7}_{-1.8}$
                                  & $^{+2.4}_{-2.3}$
                                          & &
                                                  & $^{+0.0}_{-0.1}$
                                                          & $^{+2.5}_{-2.4}$
                                                                  & $^{+4.1}_{-4.5}$
                                                                             & $^{+4.3}_{-4.6}$
                                                                                          & $^{+8.8}_{-3.2}$    \\
        ISR/FSR               & \multicolumn{2}{c}{$^{+3.0}_{+3.0}$} &       &       &       &           &           &      \\
        Cross-Section     &  5  & 5 &   &    & 10 & 10 & 6    & 30      & 30      \\
        Multijet Norm.  (shape)   &   &    & 2.5 &  &      & &          &           &           \\
        Mistag (S) & & & & $^{+36.7}_{-30}$ & & & & &\\
      \end{tabular}
\vskip 0.5cm
{\centerline{CDF $\nu\nu b\bar{b}$ loose double tag (SJ) channel relative uncertainties (\%)}}
\vskip 0.099cm
     \begin{tabular}{lccccccccc}\\
        Contribution & ZH & WH & Multijet & Mistags & Top Pair & S. Top  & Diboson  & W + HF  & Z + HF \\\hline
        Luminosity       & 3.8  & 3.8  &   &  & 3.8  & 3.8  & 3.8      & 3.8      & 3.8     \\
        Lumi Monitor      & 4.4  & 4.4  &   &  & 4.4  & 4.4  & 4.4      & 4.4      & 4.4     \\
        Tagging SF        & 8.3 & 8.3 &   &  & 8.3 & 8.3 & 8.3     & 8.3     & 8.3     \\
      Trigger Eff. (S)& 1.2 & 1.7 & 1.6 & & 0.9 & 1.8 & 2.0     & 2.5     & 1.9     \\
        Lepton Veto       & 2.0  & 2.0  &    & & 2.0  & 2.0  &2.0       & 2.0      & 2.0     \\
        PDF Acceptance    & 3.0  & 3.0  &  &   & 3.0  & 3.0  & 3.0       & 3.0      & 3.0     \\
        JES (S)       & $^{+1.9}_{-1.9}$
                                   & $^{+2.4}_{-2.4}$
                                          & &
                                                          & $^{+3.0}_{-2.8}$
                                                                        & $^{-0.6}_{0.2}$
                                                                    & $^{+4.2}_{-4.2}$
                                                                                 & $^{+6.8}_{-5.9}$
                                                                                              & $^{+8.3}_{-3.1}$    \\
        ISR/FSR               & \multicolumn{2}{c}{$^{+2.4}_{-2.4}$} &    &   &       &       &           &           &      \\
        Cross-Section     &  5.0   & 5.0 &   &   & 10 & 10 & 6    & 30      & 30      \\
        Multijet Norm.  &       & & 1.6 &       & &          &           &           \\
        Mistag (S) & & & & $^{+65.2}_{-38.5}$ & & & & &\\
      \end{tabular}
\end{ruledtabular}

\end{table}

%% file: cdf-llbb-sys.tex
\begin{table}
\begin{center}
\caption{\label{tab:cdfllbb1} Systematic uncertainties for the CDF \llbb\ 
    single tight tag (Tx) and double loose tag (LL) channels.  Systematic 
    uncertainties are listed by name; see the original references for a 
    detailed explanation of their meaning and on how they are derived.  
    Uncertainties are relative, in percent on the event yield. Shape 
    uncertainties are labeled with an ``(S)''.}
\vskip 0.2cm
{\centerline{CDF $\ell\ell b \bar{b}$ single tight tag (TT) channels relative uncertainties (\%)}}
\vskip 0.099cm
\begin{ruledtabular}
\begin{tabular}{lccccccccc} \\
Contribution   & ~Fakes~ & ~~~$t\bar{t}$~~~  & ~~$WW$~~ & ~~$WZ$~~ & ~~$ZZ$~~  & ~$Z+c{\bar{c}}$~ & ~$Z+b{\bar{b}}$~& ~Mistags~ & ~~~$ZH$~~~ \\ \hline
Luminosity ($\sigma_{\mathrm{inel}}(p{\bar{p}})$)          &     &    3.8 &     3.8&3.8&3.8 &    3.8           &    3.8          &        &    3.8  \\
Luminosity Monitor        &     &    4.4 & 4.4 & 4.4 &      4.4 &    4.4           &    4.4          &       &    4.4  \\
Lepton ID    &     &    1 &    1& 1& 1 &      1           &    1          &       &    1  \\
Lepton Energy Scale    &     &    1.5 &      1.5 & 1.5& 1.5 &    1.5           &    1.5          &        &    1.5  \\
Fake $Z\rightarrow e^+ e^-$       & 50    &   & &   &    &              &             &        &     \\
Fake $Z\rightarrow \mu^+ \mu^-$       & 5    &   & &   &    &              &             &        &     \\
Tight Mistag Rate  &   &   & & &  &  &  & 19 &  \\
Loose Mistag Rate  &   &   & & &  &  &  &  &  \\
JES  [$e^+ e^-$, 2 jet]     &     &
$^{-0.3}_{+0.3}$   &   
$^{+13.7}_{-13.5}$   &   
$^{+8.5}_{-8.5}$   &   
$^{+6.5}_{-6.3}$   &   
$^{+13.2}_{-13.2}$   &   
$^{+11.0}_{-11.1}$   &   
$^{+12.0}_{-12.0}$   &   
$^{+3.5}_{-3.8}$   \\   
JES [$e^+ e^-$, 3 jet]        &     &
$^{+7.1}_{-7.1}$   &   
$^{+8.9}_{-8.2}$   &   
$^{+17.0}_{-17.0}$   &   
$^{+15.4}_{-15.4}$   &   
$^{+16.4}_{-16.4}$   &   
$^{+15.8}_{-15.9}$   &   
$^{+18.6}_{-18.5}$   &   
$^{+15.4}_{-15.7}$   \\   
JES  [$\mu^+ \mu^-$, 2 jet]     &     &
$^{+0.6}_{-0.7}$   &   
$^{+3.9}_{-3.3}$   &   
$^{+8.6}_{-8.6}$   &   
$^{+7.6}_{-7.7}$   &   
$^{+10.2}_{-10.5}$   &   
$^{+9.3}_{-9.3}$   &   
$^{+11.1}_{-11.1}$   &   
$^{+3.4}_{-3.7}$   \\  
JES  [$\mu^+ \mu^-$, 3 jet]     &     &
$^{+5.5}_{-5.5}$   &   
$^{+5.7}_{-1.9}$   &   
$^{+16.6}_{-16.6}$   &   
$^{+16.8}_{-16.8}$   &   
$^{+16.1}_{-16.2}$   &   
$^{+16.1}_{-16.2}$   &   
$^{+17.5}_{-17.5}$   &   
$^{+13.8}_{-13.9}$   \\   
Tight $b$-tag Rate       &     &    3.9 &      3.9 & 3.9 & 3.9 &    3.9           &   3.9         &      &    3.9 \\
Loose $b$-tag Rate     &     &     &      &  &  &               &           &      &     \\
$t{\bar{t}}$ Cross Section         &     &   10&  &&      &              &             &        &     \\
Diboson Cross Section        &     &    & 6 & 6& 6    &              &             &        &     \\
$Z+$HF Cross Section      &        &  &&  &    &  40            & 40           &        &     \\
$ZH$ Cross Section    &     &    &   &&   &              &             &        &    5 \\
ISR/FSR           &     &    &     &&  &              &             &        &   0.9--12.8\\
Electron Trigger Eff.  &        & 1    & 1 & 1& 1   & 1              & 1             &      &   1     \\
Muon Trigger Eff.  &        & 5   & 5 & 5& 5   & 5              & 5             &      &   5    \\
\end{tabular}
\end{ruledtabular}
\vskip 0.5cm
{\centerline{CDF $\ell\ell b \bar{b}$ double loose tag (LL) channels relative uncertainties (\%)}}
\vskip 0.099cm
\begin{ruledtabular}
\begin{tabular}{lccccccccc} \\
Contribution   & ~Fakes~ & ~~~$t\bar{t}$~~~  & ~~$WW$~~ & ~~$WZ$~~ & ~~$ZZ$~~  & ~$Z+c{\bar{c}}$~ & ~$Z+b{\bar{b}}$~& ~Mistags~ & ~~~$ZH$~~~ \\ \hline
Luminosity ($\sigma_{\mathrm{inel}}(p{\bar{p}})$)          &     &    3.8 &     3.8&3.8&3.8 &    3.8           &    3.8          &        &    3.8  \\
Luminosity Monitor        &     &    4.4 & 4.4 & 4.4 &      4.4 &    4.4           &    4.4          &       &    4.4  \\
Lepton ID    &     &    1 &    1& 1& 1 &      1           &    1          &       &    1  \\
Lepton Energy Scale    &     &    1.5 &      1.5 & 1.5& 1.5 &    1.5           &    1.5          &        &    1.5  \\
Fake $Z\rightarrow e^+ e^-$       & 50    &   & &   &    &              &             &        &     \\
Fake $Z\rightarrow \mu^+ \mu^-$       & 5    &   & &   &    &              &             &        &     \\
Tight Mistag Rate  &   &   & & &  &  &  &  &  \\
Loose Mistag Rate  &   &   & & &  &  &  & 20 &  \\
JES  [$e^+ e^-$, 2 jet]     &     &
$^{+0.5}_{-0.5}$   &   
$^{+7.5}_{-4.8}$   &   
$^{+8.6}_{-8.7}$   &   
$^{+9.0}_{-8.9}$   &   
$^{+10.0}_{-9.3}$   &   
$^{+11.3}_{-11.0}$   &   
$^{+12.5}_{-12.5}$   &   
$^{+4.0}_{-4.4}$   \\   
JES [$e^+ e^-$, 3 jet]        &     &
$^{+8.6}_{-8.6}$   &   
$^{+32.9}_{-29.5}$   &   
$^{+14.6}_{-14.9}$   &   
$^{+16.5}_{-15.2}$   &   
$^{+20.8}_{-20.8}$   &   
$^{+17.8}_{-17.9}$   &   
$^{+18.9}_{-19.0}$   &   
$^{+14.6}_{-15.4}$   \\   
JES  [$\mu^+ \mu^-$, 2 jet]     &     &
$^{+2.5}_{-2.5}$   &   
$^{+4.5}_{-3.0}$   &   
$^{+6.7}_{-6.7}$   &   
$^{+10.2}_{-9.9}$   &   
$^{+9.2}_{-9.3}$   &   
$^{+7.7}_{-7.6}$   &   
$^{+11.5}_{-11.5}$   &   
$^{+3.9}_{-4.3}$   \\   
JES  [$\mu^+ \mu^-$, 3 jet]     &     &
$^{+9.2}_{-9.2}$   &   
$^{+13.4}_{-10.4}$   &   
$^{+14.1}_{-14.1}$   &   
$^{+16.6}_{-16.6}$   &   
$^{+14.7}_{-14.7}$   &   
$^{+16.8}_{-16.9}$   &   
$^{+17.5}_{-17.5}$   &   
$^{+11.6}_{-12.2}$   \\   
Tight $b$-tag Rate     &     &     &      &  &  &               &           &      &     \\
Loose $b$-tag Rate       &     &    6.3 &      6.3 & 6.3 & 6.3 &    6.3          &   6.3         &      &    6.3 \\
$t{\bar{t}}$ Cross Section         &     &   10&  &&      &              &             &        &     \\
Diboson Cross Section        &     &    & 6 & 6& 6    &              &             &        &     \\
$Z+$HF Cross Section      &        &  &&  &    &  40            & 40           &        &     \\
$ZH$ Cross Section    &     &    &   &&   &              &             &        &    5 \\
ISR/FSR           &     &    &     &&  &              &             &        &   3.1--15.2 \\
Electron Trigger Eff.  &        & 1    & 1 & 1& 1   & 1              & 1             &      &   1     \\
Muon Trigger Eff.  &        & 5   & 5 & 5& 5   & 5              & 5             &      &   5    \\
\end{tabular}
\end{ruledtabular}

\end{center}
\end{table}

\begin{table}
\begin{center}
\caption{\label{tab:cdfllbb2} Systematic uncertainties for the CDF \llbb\ 
    tight double tag (TT) and one tight tag and one loose tag (TL) channels. 
    Systematic uncertainties are listed by name; see the original references 
    for a detailed explanation of their meaning and on how they are derived.  
    Uncertainties are relative, in percent on the event yield. Shape 
    uncertainties are labeled with an ``(S)''.}
\vskip 0.2cm
{\centerline{CDF $\ell\ell b \bar{b}$ tight double tag (TT) channels relative uncertainties (\%)}}
\vskip 0.099cm
\begin{ruledtabular}
\begin{tabular}{lccccccccc} \\
Contribution   & ~Fakes~ & ~~~$t\bar{t}$~~~  & ~~$WW$~~ & ~~$WZ$~~ & ~~$ZZ$~~  & ~$Z+c{\bar{c}}$~ & ~$Z+b{\bar{b}}$~& ~Mistags~ & ~~~$ZH$~~~ \\ \hline
Luminosity ($\sigma_{\mathrm{inel}}(p{\bar{p}})$)          &     &    3.8 &     3.8&3.8&3.8 &    3.8           &    3.8          &        &    3.8  \\
Luminosity Monitor        &     &    4.4 & 4.4 & 4.4 &      4.4 &    4.4           &    4.4          &       &    4.4  \\
Lepton ID    &     &    1 &    1& 1& 1 &      1           &    1          &       &    1  \\
Lepton Energy Scale    &     &    1.5 &      1.5 & 1.5& 1.5 &    1.5           &    1.5          &        &    1.5  \\
Fake $Z\rightarrow e^+ e^-$       & 50    &   & &   &    &              &             &        &     \\
Fake $Z\rightarrow \mu^+ \mu^-$       & 5    &   & &   &    &              &             &        &     \\
Tight Mistag Rate  &   &   & & &  &  &  & 40 &  \\
Loose Mistag Rate  &   &   & & &  &  &  &  &  \\
JES  [$e^+ e^-$, 2 jet]     &     &
$^{+0.8}_{-0.7}$   &   
$^{+14.4}_{-13.2}$   &   
$^{+6.2}_{-6.2}$   &   
$^{+8.2}_{-8.3}$   &   
$^{+5.6}_{-5.6}$   &   
$^{+8.1}_{-7.9}$   &   
$^{+10.4}_{-10.4}$   &   
$^{+3.6}_{-4.2}$   \\  
JES [$e^+ e^-$, 3 jet]        &     &
$^{+8.3}_{-8.2}$   &   
$^{-0.7}_{+1.7}$   &   
$^{-4.2}_{+4.3}$   &   
$^{+14.4}_{-13.3}$   &   
$^{+10.6}_{-10.5}$   &   
$^{+13.2}_{-13.2}$   &   
$^{+12.4}_{-12.4}$   &   
$^{+15.1}_{-14.9}$   \\   
JES  [$\mu^+ \mu^-$, 2 jet]     &     &
$^{+1.0}_{-0.9}$   &   
$^{+5.4}_{+2.1}$   &   
$^{+13.4}_{-13.4}$   &   
$^{+7.7}_{-7.7}$   &   
$^{-1.5}_{+1.5}$   &   
$^{+8.2}_{-8.2}$   &   
$^{+5.7}_{-5.8}$   &   
$^{+3.1}_{-3.5}$   \\   
JES  [$\mu^+ \mu^-$, 3 jet]     &     &
$^{+9.3}_{-9.1}$   &   
$^{+3.9}_{-3.0}$   &   
$^{+4.8}_{-5.7}$   &   
$^{+15.5}_{-15.5}$   &   
$^{+7.3}_{-7.3}$   &   
$^{+14.2}_{-14.5}$   &   
$^{+20.5}_{-18.0}$   &   
$^{+12.5}_{-13.3}$   \\   
Tight $b$-tag Rate       &     &    7.8 &      7.8 & 7.8 & 7.8 &    7.8           &   7.8         &      &    7.8 \\
Loose $b$-tag Rate       &     &     &       &  &  &              &           &      &     \\
$t{\bar{t}}$ Cross Section         &     &   10&  &&      &              &             &        &     \\
Diboson Cross Section        &     &    & 6 & 6& 6    &              &             &        &     \\
$Z+$HF Cross Section      &        &  &&  &    &  40            & 40           &        &     \\
$ZH$ Cross Section    &     &    &   &&   &              &             &        &    5 \\
ISR/FSR           &     &    &     &&  &              &             &        &   5.5--7.6 \\
Electron Trigger Eff.  &        & 1    & 1 & 1& 1   & 1              & 1             &      &   1     \\
Muon Trigger Eff.  &        & 5   & 5 & 5& 5   & 5              & 5             &      &   5    \\
\end{tabular}
\end{ruledtabular}
\vskip 0.5cm
{\centerline{CDF $\ell\ell b \bar{b}$ one tight and one loose tag (TL) channels relative uncertainties (\%)}}
\vskip 0.099cm
\begin{ruledtabular}
\begin{tabular}{lccccccccc} \\
Contribution   & ~Fakes~ & ~~~$t\bar{t}$~~~  & ~~$WW$~~ & ~~$WZ$~~ & ~~$ZZ$~~  & ~$Z+c{\bar{c}}$~ & ~$Z+b{\bar{b}}$~& ~Mistags~ & ~~~$ZH$~~~ \\ \hline
Luminosity ($\sigma_{\mathrm{inel}}(p{\bar{p}})$)          &     &    3.8 &     3.8&3.8&3.8 &    3.8           &    3.8          &        &    3.8  \\
Luminosity Monitor        &     &    4.4 & 4.4 & 4.4 &      4.4 &    4.4           &    4.4          &       &    4.4  \\
Lepton ID    &     &    1 &    1& 1& 1 &      1           &    1          &       &    1  \\
Lepton Energy Scale    &     &    1.5 &      1.5 & 1.5& 1.5 &    1.5           &    1.5          &        &    1.5  \\
Fake $Z\rightarrow e^+ e^-$       & 50    &   & &   &    &              &             &        &     \\
Fake $Z\rightarrow \mu^+ \mu^-$       & 5    &   & &   &    &              &             &        &     \\
Tight Mistag Rate  &   &   & & &  &  &  & 19 &  \\
Loose Mistag Rate  &   &   & & &  &  &  & 10 &  \\
JES  [$e^+ e^-$, 2 jet]     &     &
$^{+0.9}_{-1.0}$   &   
$^{+13.0}_{-12.6}$   &   
$^{+9.3}_{-9.4}$   &   
$^{+10.3}_{-10.2}$   &   
$^{+10.3}_{-10.3}$   &   
$^{+8.9}_{-9.3}$   &   
$^{+10.4}_{-10.4}$   &   
$^{+4.0}_{-4.2}$   \\   
JES [$e^+ e^-$, 3 jet]        &     &
$^{+6.9}_{-7.0}$   &   
$^{+10.3}_{-8.3}$   &   
$^{+16.2}_{-16.0}$   &   
$^{+14.6}_{-14.5}$   &   
$^{+22.8}_{-23.4}$   &   
$^{+15.1}_{-15.2}$   &   
$^{+18.5}_{-18.5}$   &   
$^{+14.3}_{-14.4}$   \\   
JES  [$\mu^+ \mu^-$, 2 jet]     &     &
$^{+1.1}_{-1.1}$   &   
$^{+3.7}_{1.8}$   &   
$^{+6.5}_{-6.5}$   &   
$^{+7.5}_{-7.5}$   &   
$^{+12.5}_{-12.4}$   &   
$^{+10.1}_{-10.1}$   &   
$^{+11.0}_{-11.0}$   &   
$^{+4.0}_{-4.1}$   \\   
JES  [$\mu^+ \mu^-$, 3 jet]     &     &
$^{+8.0}_{-8.0}$   &   
$^{+2.0}_{-1.6}$   &   
$^{+14.4}_{-14.5}$   &   
$^{+24.1}_{-24.1}$   &   
$^{+16.0}_{-14.7}$   &   
$^{+17.5}_{-17.6}$   &   
$^{+14.3}_{-14.2}$   &   
$^{+13.1}_{-14.0}$   \\   
Tight $b$-tag Rate       &     &    3.9 &      3.9 & 3.9 & 3.9 &    3.9           &   3.9         &      &    3.9 \\
Loose $b$-tag Rate     &     &    3.2 &      3.2 & 3.2 & 3.2 &    3.2           &   3.2        &      &    3.2 \\
$t{\bar{t}}$ Cross Section         &     &   10&  &&      &              &             &        &     \\
Diboson Cross Section        &     &    & 6 & 6& 6    &              &             &        &     \\
$Z+$HF Cross Section      &        &  &&  &    &  40            & 40           &        &     \\
$ZH$ Cross Section    &     &    &   &&   &              &             &        &    5 \\
ISR/FSR           &     &    &     &&  &              &             &        &   3.4--7.0 \\
Electron Trigger Eff.  &        & 1    & 1 & 1& 1   & 1              & 1             &      &   1     \\
Muon Trigger Eff.  &        & 5   & 5 & 5& 5   & 5              & 5             &      &   5    \\
\end{tabular}
\end{ruledtabular}

\end{center}
\end{table}

%% file: d0-lvbb-sys.tex

\begin{table}[h]
\begin{center}
  \caption{\label{tab:d0systwh} Systematic uncertainties for the D0
    $\ell\nu b{\bar{b}}$ single tag (ST) and double tag (DT) channels.
    Systematic uncertainties are listed by name; see the original
    references for a detailed explanation of their meaning and on how
    they are derived.  Uncertainties are relative, in percent on the
    event yield.  Shape uncertainties are labeled with an ``(S)'', and
    ``SO'' represents uncetrainties that affect only the shape, but
    not the event yield.  }
\vskip 0.2cm
{\centerline{D0 $\ell\nu b\bar{b}$ Single Tag (ST) channels relative uncertainties (\%)}}
\vskip 0.099cm
\begin{ruledtabular}
\begin{tabular}{l c c c c c c }\\
Contribution                   &~Dibosons~ & $W+b\bar{b}/c\bar{c}$& $W$+l.f. & $~~~t\bar{t}~~~$ &single top&Multijet\\
\hline
Luminosity                     &  6.1    &  6.1  &    6.1  &    6.1  &    6.1  &   -- \\ 
Electron ID/Trigger efficiency   (S) & 1--5    & 2--4  &    2--4 &   1--2  &   1--2  &   -- \\       
Muon Trigger efficiency (S)          &  1--3   &  1--2 &    1--3 &    2--5 &    2--3 &   -- \\       
Muon ID efficiency/resolution       &   4.1   &   4.1 &     4.1 &     4.1 &     4.1 &   -- \\        
Jet ID efficiency  (S)          &  2--5   &  1--2 &    1--3 &    3--5 &    2--4 &   -- \\ 
Jet Energy Resolution (S)          &  4--7   &  1--3 &    1--4 &    2--5 &    2--4 &   -- \\       
Jet Energy Scale  (S)          &  4--7   &  2--5 &    2--5 &    2--5 &    2--4 &   -- \\       
Vertex Conf. Jet  (S)          & 4--10   & 5--12 &   4--10 &   7--10 &   5--10 &   -- \\       
$b$-tag/taggability (S)        & 1--4    &  1--2 &   3--7  &    3--5 &    1--2 &   -- \\ 
Heavy-Flavor K-factor          &   --    &    20 &      -- &   --    &   --    &   -- \\       
Multijet model, $e\nu b\bar{b}$ (S)   & 1--2    & 2--4  &    1--3 & 1--2    &    1--3 &   15 \\ 
Multijet model, $\mu\nu b\bar{b} $    &  --     &   2.4 &   2.4   &   --    &   --    &   20 \\ 
Cross Section                  &     6   &     9 &     9   &    10   &      10 &   -- \\ 
ALPGEN MLM pos/neg(S)          &   --    &   SO  &      -- &   --    &   --    &   -- \\       
ALPGEN Scale (S)               &   --    &   SO  &    SO   &   --    &   --    &   -- \\       
Underlying Event (S)           &   --    &   SO  &      -- &   --    &   --    &   -- \\       
PDF, reweighting               &  2      &  2    & 2       & 2       &    2    &   -- \\
\end{tabular}
\end{ruledtabular}
\vskip 0.5cm
{\centerline{D0 $\ell\nu b\bar{b}$ Double Tag (DT) channels relative uncertainties (\%)}}
\vskip 0.099cm
\begin{ruledtabular}
\begin{tabular}{ l c c c c c c }   \\
Contribution  &~Dibosons~&$W+b\bar{b}/c\bar{c}$&$W$+l.f.&$~~~t\bar{t}~~~$&single top&Multijet \\
\hline
Luminosity                    &  6.1  &  6.1  &  6.1  &  6.1  &  6.1  &   --     \\ 
Electron ID/Trigger  efficiency (S)  & 2--5  & 2--3  &  2--3 & 1--2  & 1--2  &   --     \\       
Muon Trigger efficiency (S)         &  2--4 &  1--2 &  1--2 &  2--4 &  1--3 &   --     \\       
Muon ID efficiency/resolution      &   4.1 &   4.1 &   4.1 &   4.1 &   4.1 &   --     \\        
Jet ID efficiency  (S)         &  2--8 &  2--5 &  4--9 &  3--7 &  2--4 &   --     \\ 
Jet Energy Resolution    (S)         &  4--7 &  2--7 &  2--7 &  2--9 &  2--4 &   --     \\       
Jet Energy Scale  (S)         &  4--7 &  2--6 &  2--7 &  2--6 &  2--7 &   --     \\       
Vertex Conf. Jet  (S)         & 4--10 & 5--12 & 4--10 & 7--10 & 5--10 &   --     \\       
$b$-tag/taggability (S)       & 3--7  &  4--6 & 3--10 & 5--10 & 4--10 &   --     \\ 
Heavy-Flavor K-factor         &   --  &    20 &    -- &  --   &  --   &   --     \\       
Multijet model, $e\nu\bb$ (S)  & 1--2  & 2--4  & 1--3  & 1--2  &  1--3 &   15     \\ 
Multijet model, $\mu\nu\bb$   &  --   &   2.4 &   2.4 & --    &  --   &   20     \\ 
Cross Section                 &     6 &     9 &     9 &    10 &    10 &   --     \\
ALPGEN MLM pos/neg(S)         &   --  &   SO  &    -- &   --  &   --  &   --     \\       
ALPGEN Scale (S)              &   --  &   SO  &    SO &   --  &   --  &   --     \\       
Underlying Event (S)          &   --  &   SO  &    -- &   --  &   --  &   --     \\       
PDF, reweighting              &  2    &  2    & 2     & 2     &  2    &   --     \\
\end{tabular}
\end{ruledtabular}

\end{center}
\end{table}

%% file: d0-vvbb-sys.tex
\begin{table}[h]
  \caption{\label{tab:d0systzhll} Systematic uncertainties for the D0 $\nu\nu b{\bar{b}}$
    single tag (ST) and double tag (DT) channels. Systematic uncertainties
    are listed by name; see the original references for a detailed
    explanation of their meaning and on how they are derived. Uncertainties
    are relative, in percent on the event yield.  Shape uncertainties are
    labeled with an ``(S)'', and ``SO'' represents shape only uncertainty.}
\vskip 0.2cm
{\centerline{D0 $\nu\nu b\bar{b}$ Single Tag (ST) channels relative uncertainties (\%)}}
\vskip 0.099cm
\begin{ruledtabular}
\begin{tabular}{l c c c c  c }\\
Contribution            & Top  & $V+b\bar{b}/c\bar{c}$ & $V$+l.f. & Dibosons & Multijet \\
\hline
Jet ID efficiency (S)      & 2.0  &  2.0   &  2.0   &  2.0  & -- \\
Jet Energy Scale (S)            & 2.2  &  1.6   &  3.1   &  1.0  & -- \\
Jet Energy Resolution (S)              & 0.5  &  0.3   &  0.3   &  0.9  & -- \\
Vertex Conf. / Taggability (S)  & 3.2  &  1.9   &  1.7   &  1.8  & -- \\
b Tagging (S)                   & 1.1  &  0.8   &  1.8   &  1.2  & -- \\
Lepton Identification           & 1.6  &  0.9   &  0.8   &  1.0  & -- \\
Trigger                         & 2.0  &  2.0   &  2.0   &  2.0  & -- \\
Heavy Flavor Fractions          & --   &  20.0  &  --    &  --   & -- \\
Multijet model                  & --   &  --    &  --    &  --   & 25 \\
Cross Sections                  & 10.0 &  10.2  &  10.2  &  7.0  & -- \\
Luminosity                      & 6.1  &  6.1   &  6.1   &  6.1  & -- \\
Multijet Normalilzation         & --   &  --    &  --    &  --   & -- \\
ALPGEN MLM (S)                  & --   &  --    &  SO    &  --   & -- \\
ALPGEN Scale (S)                & --   &  SO    &  SO    &  --   & -- \\
Underlying Event (S)            & --   &  SO    &  SO    &  --   & -- \\
PDF, reweighting (S)            & SO   &  SO    &  SO    &  SO   & -- \\
\end{tabular}
\vskip 0.5cm
{\centerline{D0 $\nu\nu b\bar{b}$ Double Tag (DT) channels relative uncertainties (\%)}}
\vskip 0.099cm
\begin{tabular}{ l c c c c c }   \\
Contribution            & Top  & $V+b\bar{b}/c\bar{c}$ & $V$+l.f. & Dibosons & Multijet \\
\hline
Jet ID efficiency          & 2.0  &  2.0   &  2.0   &  2.0   & -- \\
Jet Energy Scale                & 2.1  &  1.6   &  3.4   &  1.2   & -- \\
Jet Energy Resolution                  & 0.7  &  0.4   &  0.5   &  1.5   & -- \\
Vertex Conf. / Taggability      & 2.6  &  1.6   &  1.6   &  1.8   & -- \\
b Tagging                       & 6.2  &  4.3   &  4.3   &  3.7   & -- \\
Lepton Identification           & 2.0  &  0.9   &  0.8   &  0.9   & -- \\
Trigger                         & 2.0  &  2.0   &  2.0   &  2.0   & -- \\
Heavy Flavor Fractions          & --   &  20.0  &  --    &  --    & -- \\
Multijet model                  & --   &  --    &  --    &  --   & 25 \\
Cross Sections                  & 10.0 &  10.2  &  10.2  &  7.0   & -- \\
Luminosity                      & 6.1  &  6.1   &  6.1   &  6.1   & -- \\
Multijet Normalilzation         & --   &  --    &  --    &  --    & -- \\
ALPGEN MLM pos/neg (S)          &  --  &  --    &  SO    &  --    & -- \\
ALPGEN Scale (S)                &  --  &  SO    &  SO    &  --    & -- \\
Underlying Event (S)            &  --  &  SO    &  SO    &  --    & -- \\
PDF, reweighting (S)            &  SO  &  SO    &  SO    &  SO    & -- \\

\end{tabular}
\end{ruledtabular}

\end{table}

%% file: d0-llbb-sys.tex
\begin{table}
  \caption{\label{tab:d0llbb1} Systematic uncertainties for the D0 \llbb\ 
    single tag (ST) and double tag (DT) channels. Systematic uncertainties are listed by name; see the original
    references for a detailed explanation of their meaning and on how they are derived.
    Uncertainties are relative, in percent on the event yield. Shape uncertainties are
    labeled with an ``(S)''. }
\vskip 0.2cm
{\centerline{D0 $\ell\ell b \bar{b}$ Single Tag (ST) channels relative uncertainties (\%)}}
\vskip 0.099cm
\begin{ruledtabular}
\begin{tabular}{  l  c  c  c  c  c  c  c }   
Contribution               & Multijet& $Z$+l.f.  &  $Z+\bb$ & $Z+\cc$ & Dibosons & Top\\ \hline
Jet Energy Scale (S)       &   --    &  3.0   &  8.4   &  10   &  3.3   &  1.5  \\
Jet Energy Resolution (S)  &   --    &  3.9   &  5.2   &  5.3  & 0.04   &  0.6  \\
Jet ID efficiency (S)      &   --    &  0.9   &  0.6   &  0.2  &  1.0   &  0.3  \\
Taggability (S)            &   --    &  5.2   &  7.2   &  7.3  &  6.9   &  6.5  \\
$Z p_T$ Model (S)          &   --    &  2.7   &  1.4   &  1.5  &   --   &   --  \\
HF Tagging Efficiency (S)  &   --    &   --   &  5.0   &  9.4  &   --   &  5.2  \\
LF Tagging Efficiency (S)  &   --    &   73   &   --   &   --  &  5.8   &   --  \\
$ee$ Multijet Shape (S)    &   53    &   --   &   --   &   --  &   --   &   --  \\
Multijet Normalization     &  20-50  &   --   &   --   &   --  &   --   &   --  \\
$Z$+jets Jet Angles (S)    &   --    &  1.7   &  2.7   &  2.8  &   --   &   --  \\
Alpgen MLM (S)             &   --    &  0.3   &   --   &   --  &   --   &   --  \\
Alpgen Scale (S)           &   --    &  0.4   &  0.2   &  0.2  &   --   &   --  \\
Underlying Event (S)       &   --    &  0.2   &  0.1   &  0.1  &   --   &   --  \\
Trigger (S)                &   --    &  0.03  &  0.2   &  0.3  &  0.3   &  0.4  \\
Cross Sections             &   --    &   --   &  20    &  20   &  7     &  10   \\
Normalization              &   --    &  1.3   &  1.3   &  1.3  &  8.0   &  8.0  \\
PDFs                       &   --    &  1.0   &  2.4   &  1.1  &  0.7   &  5.9 
\end{tabular}
\end{ruledtabular}
\vskip 0.5cm

{\centerline{D0 $\ell\ell b \bar{b}$ Double Tag (DT) channels relative uncertainties (\%)}}
\vskip 0.099cm
\begin{ruledtabular}
\begin{tabular}{  l  c  c  c  c  c  c  c }  \\
Contribution               & Multijet& $Z$+l.f.  &  $Z+\bb$ & $Z+\cc$ & Dibosons & Top\\  \hline
Jet Energy Scale (S)       &   --    &  4.0   &  6.4   &  8.2   &  3.8   &  2.7  \\
Jet Energy Resolution(S)   &   --    &  2.6   &  3.9   &  4.1   &  0.9   &  1.5  \\
JET ID efficiency (S)      &   --    &  0.7   &  0.3   &  0.2   &  0.7   &  0.4  \\
Taggability (S)            &   --    &  8.6   &  6.5   &  8.2   &  4.6   &  2.1  \\
$Z_{p_T}$ Model (S)        &   --    &  1.6   &  1.3   &  1.4   &   --   &   --  \\
HF Tagging Efficiency (S)  &   --    &   --   &  1.3   &  3.2   &   --   &  0.7  \\
LF Tagging Efficiency (S)  &   --    &   72   &   --   &   --   &  4.0   &   --  \\
$ee$ Multijet Shape (S)    &    59   &   --   &   --   &   --   &   --   &   --  \\
Multijet Normalization     &  20-50  &   --   &   --   &   --   &   --   &   --  \\
$Z$+jets Jet Angles (S)    &   --    &  2.0   &  1.5   &  1.5   &   --   &   --  \\
Alpgen MLM (S)             &   --    &  0.4   &   --   &   --   &   --   &   --  \\
Alpgen Scale (S)           &   --    &  0.2   &  0.2   &  0.2   &   --   &   --  \\
Underlying Event(S)        &   --    &  0.1   & 0.02   &  0.1   &   --   &   --  \\
Trigger (S)                &   --    &  0.3   &  0.2   &  0.1   &  0.2   &  0.5  \\
Cross Sections             &   --    &   --   & 20     & 20     & 7      & 10    \\
Normalization              &   --    &  1.3   & 1.3    & 1.3    & 8.0    & 8.0   \\
PDFs                       &   --    &  1.0   & 2.4    & 1.1    & 0.7    & 5.9 
\end{tabular}
\end{ruledtabular}
\end{table}

%% file: vz_xbb_combo_note.bbl
\clearpage

\begin{thebibliography}{99}

\bibitem{dibo}  J.~M.~Campbell and R.~K.~Ellis,
Phys.\ Rev.\  D {\bf 60}, 113006 (1999).
We used {\sc MCFM} v6.0. Cross sections are computed
using a choice of scale $\mu_0^2=M_V^2+p_T^2(V)$, where $V$ is the vector boson,
and the MSTW2008 PDF set.

\bibitem{bib:anocoups} K.~Hagiwara, S.~Ishihara, R.~Szalapski, and D.~Zeppenfeld,  Phys.\ Rev.\ D {\bf 48} (1993).

\bibitem{bib:newphen} J.~C.~Pati and A.~Salam, Phys.\ Rev.\ D {\bf 10}, 275 (1974); {\bf 11} 703(E) (1975);\\
                      G.~Altarelli, B.~Mele, and M.~Ruiz-Altaba, Z.\ Phys.\ C {\bf 45}, 109 (1989); {\bf 47}, 676(E) (1990);\\
                      L.~Randall and R.~Sundrum, Phys.\ Rev.\ Lett. {\bf 83}, 3370 (1999);\\
                      H.~Davoudiasl, J.~L.~Hewett, and T.~G.~Rizzo, Phys.\ Rev.\ D {\bf 63}, 075004 (2001);\\
                      H.~He {\sl et al.}, Phys.\ Rev.\ D {\bf 78}, 031701 (2008).

\bibitem{bib:leptonic}   
  		      T.~Aaltonen  {\sl et al.} (CDF Collaboration), Phys.\ Rev.\ Lett. {\bf 104}, 201801 (2010);\\
		      V.~M.~Abazov {\sl et al.} (D0 Collaboration), Phys.\ Rev.\ Lett. {\bf 101}, 171803 (2008);\\
                         V.~M.~Abazov {\sl et al.} (D0 Collaboration), Phys.\ Lett.\ B {\bf 695}, 67 (2011);\\
                         V.~M.~Abazov {\sl et al.} (D0 Collaboration), Phys.\ Rev.\ D {\bf 84}, 011103 (2011);\\
                         V.~M.~Abazov {\sl et al.} (D0 Collaboration), arXiv:1201.5652 [hep-ex].
\bibitem{bib:hadronic}   T.~Aaltonen {\sl et al.} (CDF Collaboration), Phys.\ Rev.\ Lett. {\bf 103}, 091803 (2009);\\
                         T.~Aaltonen {\sl et al.} (CDF Collaboration), Phys.\ Rev.\ Lett. {\bf 104}, 101801 (2010);\\ 
                         V.~M.~Abazov {\sl et al.} (D0 Collaboration), arXiv:1112.0536 [hep-ex].

\bibitem{dzDibosonCombo} V.~M.~Abazov {\sl et al.} (D0 Collaboration), \DZ Note 6260-CONF (2011).

\bibitem{cdfDibosonCombo} T.~Aaltonen  {\sl et al.} (CDF Collaboration), CDF Conference Note 10805 (2012).

\bibitem{bib:higgs}      V.~M.~Abazov {\sl et al.} (D0 Collaboration), Phys.\ Rev.\ Lett. {\bf 104}, 071801 (2010);\\
                         V.~M.~Abazov {\sl et al.} (D0 Collaboration), Phys.\ Rev.\ Lett. {\bf 105}, 251801 (2010);\\
                         V.~M.~Abazov {\sl et al.} (D0 Collaboration), Phys.\ Lett. B {\bf 698}, 6 (2011); \\
                         T.~Aaltonen {\sl et al.}  (CDF Collaboration),  Phys.\ Rev.\ Lett. {\bf 105}, 251802 (2010);\\
                         T.~Aaltonen {\sl et al.}  (CDF Collaboration),  Phys.\ Rev.\ Lett. {\bf 103}, 101802 (2009);\\
                         T.~Aaltonen {\sl et al.}  (CDF Collaboration),  Phys.\ Rev.\ Lett. {\bf 104}, 141801 (2010).


\bibitem{cdfWHl} T.~Aaltonen  {\sl et al.} (CDF Collaboration), CDF Conference Note 10796 (2012).


\bibitem{cdfZHv}  T.~Aaltonen  {\sl et al.} (CDF Collaboration), CDF Conference Note 10798 (2012).

\bibitem{cdfZHl} T.~Aaltonen  {\sl et al.} (CDF Collaboration), CDF Conference Note 10799 (2012).

\bibitem{dzWHl} V.~M.~Abazov {\sl et al.} (D0 Collaboration), \DZ Note 6220-CONF (2011).

\bibitem{dzZHv} V.~M.~Abazov {\sl et al.} (D0 Collaboration), \DZ Note 6223-CONF (2011).

\bibitem{dzZHl} V.~M.~Abazov {\sl et al.} (D0 Collaboration), \DZ Note 6256-CONF (2011).


\bibitem{cdf} 
  D.~Acosta {\it et al.}  (CDF Collaboration),
  Phys.\ Rev.\ D {\bf 71}, 032001 (2005).


\bibitem{dzero} 
V.~M.~Abazov {\sl et al.} (D0 Collaboration), Nucl. Instrum. Methods Phys. Res. A {\bf 565}, 463  (2006);\\
M.~Abolins {\sl et al.}, Nucl. Instrum. Methods Phys. Res. A {\bf 584}, 75 (2008);\\
R.~Angstadt {\sl et al.}, Nucl. Instrum. Methods Phys. Res. A {\bf 622}, 298 (2010).


\bibitem{bib:btagnn} 
V.~M.~Abazov {\sl et al.} (D0 Collaboration), 
Nucl. Instrum. Methods Phys. Res. A {\bf 620}, 490 (2010).

\bibitem{bib:HOBIT} T.~Aaltonen  {\sl et al.} (CDF Collaboration), CDF Conference Note 10803 (2012).

\bibitem{bib:SecVtx} 
  D.~Acosta {\it et al.}  (CDF Collaboration),
  Phys.\ Rev.\ D {\bf 71}, 052003 (2005).


\bibitem{bib:JetProb} 
  A.~Abulencia {\it et al.}  (CDF and CDF - Run II Collaborations),
  Phys.\ Rev.\ D {\bf 74}, 072006 (2006).



\bibitem{alpgen}
M.~L.~Mangano, M.~Moretti, F.~Piccinini, R.~Pittau and A.~D.~Polosa,
J. High Energy Phys. {\bf 07}, 001 (2003).


\bibitem{singletop}
CompHEP, E.~Boos {\sl et al.}, Nucl. Instrum. Methods Phys. Res. A {\bf 534}, 250 (2004);\\
E.~Boos, V.~Bunichev, L.~Dudko, V.~Savrin, and A.~Sherstnev, Phys. Atom. Nucl. {\bf 69}, 1317 (2006).

\bibitem{pythia}
T.~Sj\"ostrand, L.~Lonnblad and S.~Mrenna,
  arXiv:hep-ph/0108264.

\bibitem{mcfm} J.~M.~Campbell and R.~K.~Ellis,
 http://mcfm.fnal.gov/. \\
  J.~M.~Campbell, R.~K.~Ellis,
  Nucl.\ Phys.\ Proc.\ Suppl.\  {\bf 205-206}, 10 (2010).

\bibitem{ttbar_xsec}    U.~Langenfeld, S.~Moch and P.~Uwer,
  Phys.\ Rev.\  D {\bf 80}, 054009 (2009).

\bibitem{schan_top_xsec}  N.~Kidonakis,
  arXiv:1005.3330 [hep-ph] (2010);\\
  N.~Kidonakis, Phys.\ Rev.\  D {\bf 81}, 054028 (2010).

\bibitem{tchan_top_xsec} N.~Kidonakis,
  Phys.\ Rev.\  D {\bf 74}, 114012 (2006).

\bibitem{geant} R.~Brun, R.~Hagelberg, M.~Hansroul, and J.~C.~Lasalle,
{\it GEANT: Simulation Program for Particle Physics Experiments.  User Guide
and Reference Manual}, CERN-DD-78-2-REV; \\
S. Agostinelli {\it et al.}, Nucl. Instrum. Methods A {\bf 506}, 250 (2003).

\bibitem{lumi}
  T.~Andeen {\sl et al.},
  FERMILAB-TM-2365 (2007).

\bibitem{dyxsec} R.~Hamberg, W.L.~van~Neerven and W.B.~Kilgore, 
                 Nucl Phys. B {\bf 359}, 343 (1991) [Erratum-ibid. B {\bf 644}, 403 (2002)].


\bibitem{pdgstats}
T. Junk, Nucl. Instrum. Methods Phys. Res. A {\bf 434}, 435 (1999); \\
A.~L.~Read,
J.\ Phys.\ G {\bf 28}, 2693 (2002).

\bibitem{pflh} W.~Fisher,
FERMILAB-TM-2386-E (2006).




\end{thebibliography}
